\documentclass{aa} 
\usepackage{graphicx}
\usepackage{gensymb}
\usepackage{txfonts}
\usepackage[hidelinks]{hyperref}
\hypersetup{
    colorlinks=true,
    allcolors=blue,}

\usepackage{orcidlink}

\begin{document} 

\DeclareRobustCommand{\orcid}[1]{\orcidlink{#1}}

\defcitealias{lamarcaDustPowerUnravelling2024a}{LM24}

\title{
The major merger--active galactic nucleus connection up to the cosmic noon
}

\subtitle{}

\author{
    A.~La~Marca
    \orcid{0000-0002-7217-5120}
    \inst{1,2,3}\thanks{\email{antonio.la.marca.astro@gmail.com}}
    \and
    B.~Margalef-Bentabol
    \orcid{0000-0001-8702-7019}
    \inst{1}
    \and
    L.~Wang \orcid{0000-0002-6736-9158}
    \inst{1, 2}
    \and
    S.~C.~Trager \orcid{0000-0001-6994-3566}
    \inst{2}
    \and V.~Rodriguez-Gomez
    \inst{4}
    \and G.~Martin
    \inst{5}
}

\institute{
    SRON Netherlands Institute for Space Research, Landleven 12, 9747 AD Groningen, The Netherlands
    \and
    Kapteyn Astronomical Institute, University of Groningen, Postbus 800, 9700 AV Groningen, The Netherlands
    \and
    European Space Agency/ESTEC, Keplerlaan 1, 2201 AZ Noordwijk, The Netherlands
    \and
    Instituto de Radioastronomía y Astrofísica, Universidad Nacional Autónoma de México, A.P. 72-3, 58089 Morelia, Mexico
    \and
    School of Physics and Astronomy, University of Nottingham, University Park, Nottingham NG7 2RD, UK
}

  \date{Received -; accepted -}

% \abstract{}{}{}{}{} 
% 5 {} token are mandatory
 
 \abstract
  {
  Galaxy major mergers are a potential mechanism for triggering active galactic nuclei (AGN) activity, but their role remains debated, particularly beyond the local Universe. We aim to shed light on the merger--AGN connection at $z=0.5$--$2$, exploiting the multi-wavelength datasets and {\it James Webb Space Telescope} (JWST) observations in the COSMOS field.
  We construct a stellar mass-limited sample and identify AGN via mid-infrared (MIR) colours, X-ray detections, and spectral energy distribution (SED) fitting.
  % 
  % We construct a stellar mass-limited sample and perform detailed spectral energy distribution (SED) modelling of these galaxies. We identify AGN in three different ways, using mid-infrared (MIR) colours, X-ray detections, and SED fitting. 
  We train convolutional neural networks to identify mergers with mock JWST observations. 
  %After analysing the properties of AGN host galaxies, 
  We create non-AGN and non-merger control samples matching the redshift, stellar mass, and star-formation rate distributions of the AGN and mergers. 
  We find AGN to be moderately more frequent in mergers than in non-mergers, with excess ratios ranging from $\sim2.5$ (X-ray AGN) to $\sim1.3$ (MIR) and $\sim 1.1$--1.2 (SED AGN). Similarly, AGN galaxies show a higher merger fraction ($f_{\rm merg}$) than non-AGN controls.
  We then study $f_{\rm merg}$ as a function of relative and absolute AGN power, utilising the AGN fraction ($f_{\rm AGN}$) and accretion disc luminosity (L$_{disc}$) parameters. We uncover a $f_{\rm merg}$--$f_{\rm AGN}$ relation with two regimes: $f_{\rm merg}$ stays roughly flat for less-dominant AGN ($f_{\rm AGN}<0.8$) but increases at $f_{\rm AGN}>0.8$ for the MIR and X-ray AGN, and more gently for SED AGN, where mergers appear to be the main triggering mechanism. Additionally, $f_{\rm merg}$ increases monotonically as a function of L$_{disc}$, for all AGN types, reaching $f_{\rm merg}>50\%$ for the most luminous AGN (L$_{disc} \gtrsim 10^{46}\,{\rm erg\,s^{-1}}$).
  Overall, our results suggest that major mergers can trigger AGN out to the cosmic noon at $z\sim2$. Furthermore, the role of major mergers shows a clear dependence on AGN luminosity and remains the principal mechanism for fuelling the most powerful AGN.
  %Overall, our results suggest that mergers play a more important role at lower redshifts than at intermediate redshifts: probably due to the fewer gas supplies available in low-$z$ galaxies, external resources are necessary to fuel nuclear activity.
  % Overall, when compared to lower redshift results, our results suggest that mergers play a less important role at intermediate redshifts, probably due to the larger gas supplies available, which may sustain AGN activity without the need for external resources. 
  %However, major mergers remain the principal mechanism fuelling the most powerful AGN, even at cosmic noon.
  }

  \keywords{Galaxies: interactions -- Galaxies: active -- Galaxies: evolution -- Techniques: image processing}

  \maketitle
%
%--------------------------------------------------------------------

\section{Introduction}

% Role of the galaxy mergers in shaping the galaxy population as we know them today. 

Under hierarchical structure formation, mergers play a crucial role in galaxy evolution.
%, transforming galaxy morphology/structure and influencing other intrinsic properties. 
Mergers can contribute significantly to the stellar-mass assembly process as two or more galaxies collide and coalesce into a single, more massive galaxy \citep[for a review see][]{somervillePhysicalModelsGalaxy2015}. During these interactions, gravity pulls and distorts the galaxies involved, changing their dynamics and morphology \citep{toomreGalacticBridgesTails1972, conseliceEarlyRapidMerging2006}. In addition, while rearranging the star and gas distributions, mergers have been observed to enhance star-formation rates \citep[SFRs;][]{martinRoleMergersInteractions2021, bickleyStarFormationCharacteristics2022}, to extreme levels in some cases \citep[][]{mihosUltraluminousStarburstsMajor1994, cibinelEarlyLatestageMergers2019}.
% Mergers and AGN triggering. The situation at higher redshift.
Mergers are also acknowledged as a viable mechanism for funnelling gas toward super-massive black holes \citep[SMBHs,][]{hopkinsUnifiedMergerdrivenModel2006, blumenthalGoFlowUnderstanding2018}. Many simulations suggest that mergers can fuel accretion onto SMBHs, triggering active galactic nuclei \citep[AGN;][]{dimatteoEnergyInputQuasars2005, blechaPowerInfraredAGN2018}, although others predict that mergers may be only a secondary path \citep{martinNormalBlackHoles2018, bhowmickSupermassiveBlackHole2020,byrne-mamahitInteractingGalaxiesIllustrisTNG2023}. Observational studies also find seemingly contradictory results. In the local Universe, some works support the scenario in which mergers trigger AGN \citep{urrutiaEvidenceQuasarActivity2008, hwangActivityGalacticNuclei2012, lacknerLateStageGalaxyMergers2014, gouldingGalaxyInteractionsTrigger2018, gaoMergersTriggerActive2020, pierceAGNTriggeringMechanisms2022}, while others reject this picture \citep{reichardLopsidednessPresentDayGalaxies2009, cisternasBulkBlackHole2011, sabaterTriggeringOpticalAGN2015, smethurstEvidenceNonmergerCoevolution2023}. Higher-redshift studies, limited to much smaller samples, find similarly mixed results \citep{allevatoXMMNewtonWideField2011, kocevskiCANDELSConstrainingAGNMerger2012, kocevskiAreComptonthickAGNs2015, mechtleyMostMassiveBlack2016, fanMostLuminousHeavily2016, marianMajorMergersAre2019, silvaGalaxyMergers252021, bonaventuraRelationAGNHostgalaxy2025}. 
Furthermore, some investigations reveal dependence on AGN luminosity and dust obscuration, indicating that mergers may be more important in triggering more luminous or more dust-obscured AGN \citep{treisterMajorGalaxyMergers2012, glikmanMajorMergersHost2015, ricciGrowingSupermassiveBlack2017, ricciHardXrayView2021, donleyEvidenceMergerdrivenGrowth2018, weigelFractionAGNsMajor2018, ellisonDefinitiveMergerAGNConnection2019, bickleyAGNsPostmergersUltraviolet2023, euclidcollaborationEuclidQuickData2025}, although not all studies confirm these findings \citep[][]{villforthHostGalaxiesLuminous2017, hewlettRedshiftEvolutionMajor2017}.
%Furthermore, other investigations revealed a dependence on AGN luminosity, indicating that mergers may be more important in triggering more luminous AGN \citep{urrutia_evidence_2008, treister_major_2012, glikman_major_2015, weigel_fraction_2018, ellison_definitive_2019, pierce_agn_2022, bickley_agns_2023}, although not all studies confirmed that \citep[e.g.][]{villforth_host_2017, hewlett_redshift_2017}. Finally, some studies found a strong connection between interacting galaxies and the dust-obscured phase of AGN \citep{ricci_growing_2017, ricci_hard_2021}.

% The merger identification. 

The first issue in investigating the merger--AGN connection is the identification of mergers.  Multiple methods have been used.
%, with differing advantages and disadvantages. 
Visual classification \citep{dargGalaxyZooProperties2010, tanakaGalaxyCruiseDeep2023} is hard to reproduce, time-consuming, and suffers from low accuracy and incompleteness \citep{huertas-companyCatalogVisuallikeMorphologies2015}. The close-pair method provides an objective selection of pre-mergers \citep[][]{knapenInteractingGalaxiesNearby2015, daviesGalaxyMassAssembly2015} but misses post-mergers. 
%Moreover, this method requires highly complete spectroscopic data and so is observationally expensive.
Non-parametric morphological statistics are reproducible and relatively fast \citep[][]{conseliceRelationshipStellarLight2003, lotzNewNonparametricApproach2004, pawlikShapeAsymmetryMorphological2016}. Nevertheless, they also suffer from severe contamination \citep{huertas-companyCatalogVisuallikeMorphologies2015}. Other studies use machine learning (ML) techniques to combine several morphological parameters \citep{nevinAccurateIdentificationGalaxy2019, snyderAutomatedDistantGalaxy2019, guzman-ortegaMorphologicalSignaturesMergers2023, hernandez-toledoSDSSIVMaNGAIncidence2023}. Deep Learning (DL) techniques are also reproducible and quick to run once trained \citep[][]{ackermannUsingTransferLearning2018, wangConsistentFrameworkComparing2020, bickleyConvolutionalNeuralNetwork2021, ciprijanovicDeepMergeClassifyingHighredshift2020, ciprijanovicDeepMergeIIBuilding2021}. However, their performance depends on the task assigned and is fundamentally limited by the quality of the training data \citep[for a review, see][]{margalef-bentabolGalaxyMergerChallenge2024}.
% The AGN selection methods. 
The second issue is the selection of AGN, which depends on complex multi-wavelength phenomena. The material accreting onto an SMBH emits radiation from different components (disk, dusty torus, jet, emission line regions), from radio to X-ray \citep[for a review, see][]{alexanderWhatDrivesGrowth2012}. Yet, not all AGN present the same signatures of current activity. %Therefore, selecting AGN with different techniques leads to AGN and host galaxies with different characteristics. 
Typically, AGN can be selected using mid-infrared (MIR) colours \citep[][]{Donely2007SpitzerPowerLaw, sternMidinfraredSelectionActive2012}, X-ray detection \citep[][]{kossMergingClusteringSwift2010}, optical emission line ratios and radio observations \citep[][]{ellisonGalaxyPairsSloan2015, gordonEffectMinorMajor2019}. The impact of different selections on the possible merger--AGN connection was demonstrated, for example, in \citet{satyapalGalaxyPairsSloan2014} and \citet[][hereafter \citetalias{lamarcaDustPowerUnravelling2024a}]{lamarcaDustPowerUnravelling2024a} which showed different AGN excesses in mergers compared with non-mergers, depending on the diagnostic used. 

% The host galaxies of the different AGN. 
It is widely accepted that SMBHs and their host galaxies are intricately connected \citep{kormendyCoevolutionNotSupermassive2013}. There is also a broad consensus that AGN identified through different diagnostics are associated with distinct host-galaxy properties \citep{heckmanCoevolutionGalaxiesSupermassive2014, hickoxObscuredActiveGalactic2018}. Radiative radio-quiet AGN mostly reside in host galaxies with typical stellar mass (M$_{\star}$) of $10^{10-11}$ M$_{\odot}$ and SFRs that are typical of star-forming galaxies. Radiative radio-loud AGN usually inhabit more massive early-type galaxies with less ongoing star formation. Jet-mode AGN reside in massive or very massive early-type galaxies with little or no star formation. X-ray-selected AGN populate slightly more massive galaxies \citep{bongiornoAccretingSupermassiveBlack2012, mountrichasComparisonStarFormation2022} than MIR-selected AGN \citep{azadiMOSDEFSurveyAGN2017, bornanciniPropertiesIRselectedActive2022}, whose hosts are usually more massive than optically selected type II AGN \citep{vietriTypeIIAGNhost2022}. Several works found evidence for a correlation between AGN activity and star formation, which depends on  AGN type. MIR AGN are commonly hosted by galaxies on or above the star-forming main sequence \citep[MS;][]{ellisonStarFormationRates2016, azadiMOSDEFSurveyAGN2017}. X-ray AGN are found both in quenching \citep[``green valley'';][]{silvermanEvolutionAGNHost2008, mullaneyALMAHerschelReveal2015, azadiPRIMUSRelationshipStar2015, cristelloInvestigatingStarFormation2024a} and star-forming galaxies \citep[][]{airdPRIMUSDependenceAGN2012, rosarioMeanStarFormation2012, santiniEnhancedStarFormation2012, mountrichasGalaxyPropertiesType2021}.  

% What we learned from the GAMA 09 paper. Is it the same at higher redshift? 
% What this paper aims to study in the Cosmic Evolution Survey (COSMOS; Scoville et al. 2007).
 
In \citetalias{lamarcaDustPowerUnravelling2024a}, we investigated the merger--AGN connection at $z\lesssim0.8$, using a multi-wavelength dataset that allowed us to analyse different AGN types over a wide range of luminosities. We measured the AGN fractional contribution to the total galaxy light ($f_{\rm AGN}$) through spectral energy distribution (SED) fitting, and reported a merger fraction--$f_{\rm AGN}$ relation with two distinct regimes: a flat merger fraction trend for relatively weaker AGN ($f_{\rm AGN}<80\%$), and a steep increase for dominant AGN ($f_{\rm AGN}\geq 80\%$). However, it is unclear whether such trends hold at higher redshifts. 
This paper aims to address this question by expanding our analysis to $z=2$ in the Cosmic Evolution Survey field \citep[COSMOS;][]{scovilleCosmicEvolutionSurvey2007}, using the recently released {\it James Webb Space Telescope} (JWST) COSMOS-Web images with high sensitivity and spatial resolution and the COSMOS2020 catalogue \citep{weaverCOSMOS2020PanchromaticView2022}. Furthermore, in \citetalias{lamarcaDustPowerUnravelling2024a}, we created mass- and redshift-matched control samples of non-mergers and non-AGN. Now, thanks to more extensive multi-wavelength coverage, we can construct better controls by also considering SFRs. 

This paper is organised as follows. In Sect. \ref{sect:data}, we explain the sample selection from the COSMOS2020 catalogue and the associated multi-wavelength data.
In Sect.~\ref{sect:Methods}, first we introduce the SED fitting tool used to derive galaxy properties and characterise the AGN contribution fraction ($f_{\rm AGN}$) to the total observed light. Then, we describe our AGN selection techniques, including MIR colours, X-ray detections, and SED fitting. Finally, we present the DL model used to identify mergers, which is trained on mock observations generated from cosmological simulations. %\citep[IllustrisTNG and Horizon-AGN;][]{pillepichFirstResultsIllustrisTNG2018, duboisDancingDarkGalactic2014}. 
%We utilise an SED fitting tool to derive galaxy properties and characterise the AGN contribution fraction ($f_{\rm AGN}$) to the total observed light. We describe this process, the AGN diagnostics, and the DL model trained on mock observations in Sect.~\ref{sect:Methods}. 
In Sect. \ref{sect:Results}, we present a detailed comparison of AGN and non-AGN host galaxies. After that, we investigate the merger--AGN connection using both a binary AGN/non-AGN classification and continuous AGN parameters (i.e. $f_{\rm AGN}$ and AGN luminosity). In Sect.~\ref{sect:Conclusions}, we present our main findings. Throughout the paper, we assume a flat $\Lambda$CDM universe with $\Omega_M=0.28$, $\Omega_{\Lambda}=0.71$, and $H_0=69.32$ km s$^{-1}$ Mpc$^{-1}$ \citep{hinshawNineyearWilkinsonMicrowave2013}.

%--------------------------------------------------------------------
\section{Data}\label{sect:data}

In this section, we first explain how we constructed our parent sample from the COSMOS2020 catalogue. Then, we introduce the JWST imaging data used to identify merging and non-merging galaxies.

\subsection{Parent sample selection}\label{sect:sample}

The Hubble Space Telescope (HST) extensively observed with its Advanced Camera for Surveys (ACS) Wide Field Channel (WFC) F814W the COSMOS field \citep{koekemoerCOSMOSSurveyHubble2007}. The HST observations are integrated with ground-based broad- and narrow-band observations, resulting in the COSMOS2020 \citep{weaverCOSMOS2020PanchromaticView2022} photometric catalogue, which spans from ultraviolet (UV) to the MIR. Measurements in the near-UV and far-UV come from the COSMOS Galaxy Evolution Explorer (GALEX) catalogue \citep{zamojskiDeepGALEXImaging2007}. Additional optical data are taken from the Canada-France-Hawaii Telescope Large Area $U$-band Deep Survey \citep[CLAUDS;][]{sawickiCFHTLargeArea2019} and the second public data release of the Hyper Suprime-Cam Subaru Strategic Program \citep[PDR2 HSC-SSP;][]{aiharaSecondDataRelease2019}. Moreover, the Subaru Suprime-Cam provided imaging in 7 broad and 12 medium bands \citep{taniguchiCosmicEvolutionSurvey2007, taniguchiSubaruCOSMOS202015}. The catalogue includes the near-IR broad and narrow bands data from the UltraVISTA survey \citep[data release 4;][]{mccrackenUltraVISTANewUltradeep2012} and the MIR data from the four {\it Spitzer}/IRAC channels from the Cosmic Dawn Survey \citep{euclidcollaborationEuclidPreparationXVII2022}.

The COSMOS2020 team performed source detection on a multi-band ``chi-squared'' $izYJHK_S$ detection image using two different methods.
% The first is a "Classic" approach, already used for the previous version \citep[COSMOS2015;][]{laigleCOSMOS2015CatalogExploring2016}, utilising \texttt{SExtractor} \citep{bertinSExtractorSoftwareSource1996}, 
In this work, we made use of the \texttt{Farmer} version, which utilises the \texttt{SEP} code \citep{barbarySEPSourceExtractor2016}. The \texttt{Farmer} catalogue photometry was made using the \texttt{Tractor} code \citep{langTractorProbabilisticAstronomical2016}. 
Two different photometric redshifts (photo$-z$) are listed, one computed using \texttt{LePhare} \citep{arnoutsMeasuringRedshiftEvolution2002, ilbertAccuratePhotometricRedshifts2006} and another using \texttt{eazy} \citep{brammerEAZYFastPublic2008}. When spectroscopic redshifts are unavailable, we adopt the photo$-z$ computed using \texttt{LePhare}, given its better performance \citep[][]{weaverCOSMOS2020PanchromaticView2022}. Overall, the photo$-z$ precision, given by the normalised median absolute deviation, is around $0.01\times(1+z)$ at $i<24.0$ mag, and $0.03\times(1+z)$ at $24.0< i <27.0$ mag. 

Following a recommendation by the COSMOS team, we set \texttt{flag\_combined} $= 0$ to avoid areas affected by bright stars. We also select sources that have the star/galaxy separation flag \texttt{lp\_type} $= 1$ (galaxy) or $=2$ (X-ray source). In addition, we focused our analysis in the redshift range  $0.5\leq z \leq 2$ for two primary reasons. First, reliable and detailed morphological classification requires a sufficient number of pixels. Second, at $z>2$, the JWST F150W filter probes the rest-frame UV emission, which mainly traces star-forming regions (usually clumpy and disturbed). In extreme cases, different star-forming regions in the same galaxy may appear as separate systems in the UV. By restricting our study to $z \leq 2$, we ensure that the data capture the rest-frame optical light, which correlates better with stellar mass and overall structure. Finally, the SED fitting procedure we employed provides qualitatively better results if data are available in the wavelength range where the AGN fraction will be measured. Because we are interested in the AGN fraction in the MIR regime, we required all sources to have at least two detections in the four IRAC channels with signal-to-noise ratio ($S/N$) $>5$. 
Table \ref{tab:counts} reports the number of detections in each survey after applying the MIR and redshift selections.

\begin{table}[]
    \centering
    \caption{Total number of galaxies in the range $0.5\leq z \leq 2$ from each survey, after the MIR selection.}
    \small
    \begin{tabular}{lr}
    \hline \hline\\[-7pt]
    Survey & Nr. galaxies \\
    \hline\\[-7pt]
    COSMOS2020 & 62\,846 \\
    X-ray & 270 \\
    %Far-IR & \\
    {\it Spitzer}/MIPS & 13\,924 \\
    {\it Herschel}/PACS & 4\,392 \\
    {\it Herschel}/SPIRE & 1\,741 \\
    \hline
    \end{tabular}
    \label{tab:counts}
\end{table}

\subsection{JWST COSMOS-Web}\label{sect:observ}

COSMOS-Web \citep[][PIs: Kartaltepe \& Casey, ID=1727]{caseyCOSMOSWebOverviewJWST2023} is a 255-hour JWST Treasury Program observing the central area of the COSMOS field. The program will cover a contiguous area of 0.54 deg$^2$ with four NIRCam imaging filters (F115W, F150W, F277W, and F444W) and target a 0.19 deg$^2$ area with MIRI F770W imaging data.
% with the F770W filter is targeted in parallel with the NIRCam observations. 
% By June 2023, roughly half of the entire field had been successfully observed (0.28 deg$^2$). 

\begin{figure}
    \centering
    \includegraphics[width=.45\textwidth]{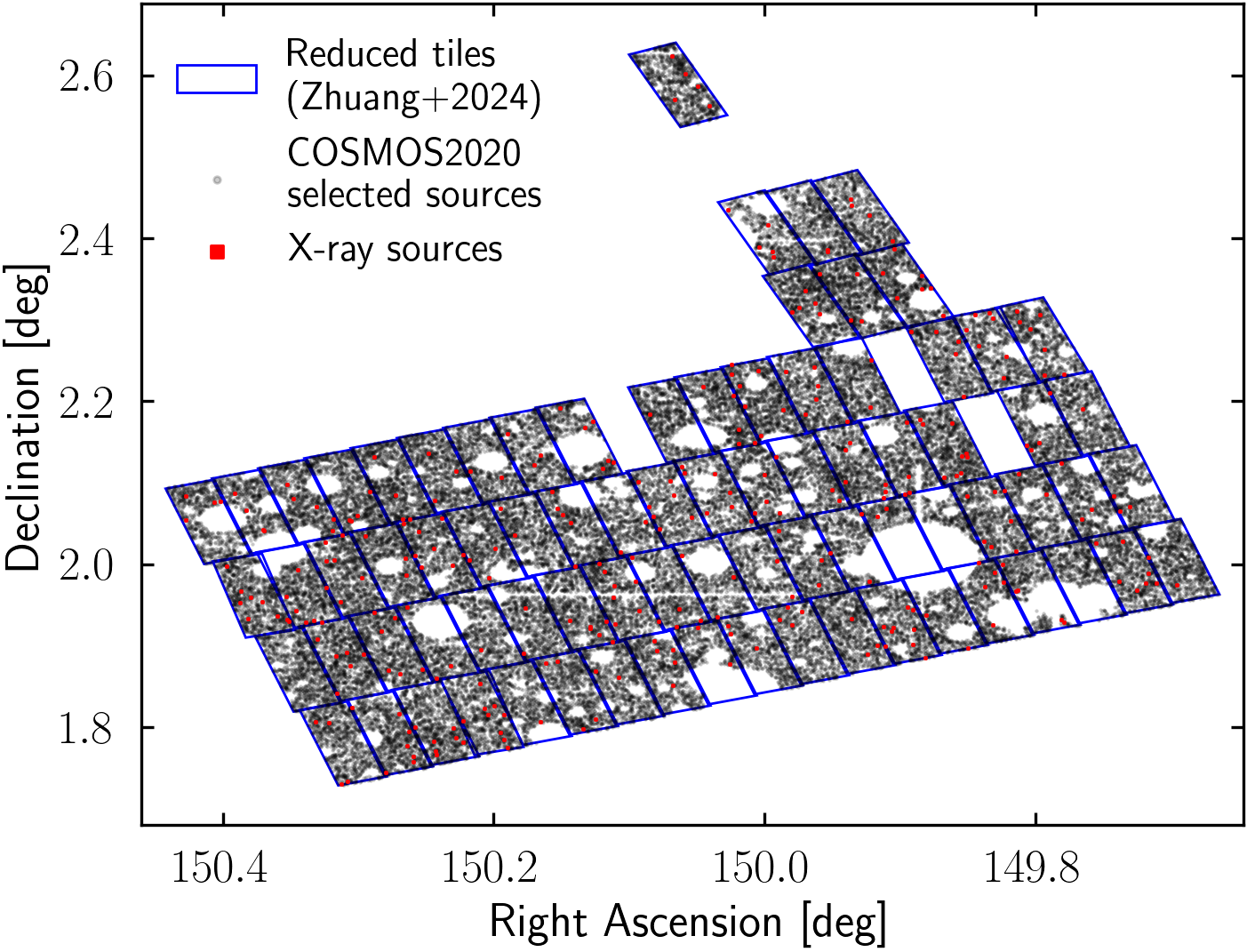}
    \caption{COSMOS2020 sources observed by the COSMOS-Web program. Blue rectangles indicate the tiles reduced by \citet{zhuangActiveGalacticNuclei2024}. Red squares are the X-ray sources identified by \citet{marchesiChandraCOSMOSLegacy2016}.}
    \label{fig:COSMOS-Web}
\end{figure}

We made use of the JWST/NIRCam F150W images over $0.28\,{\rm deg^2}$ reduced by \citet{zhuangActiveGalacticNuclei2024}, using version 1.10.2 of the \texttt{jwst}\footnote{\url{https://jwst-pipeline.readthedocs.io/en/latest/}} pipeline with the Calibration Reference Data System (CRDS) version of 11.17.0. In addition, they adopted some custom steps for the NIRCam image reduction. Briefly, they carefully treated the ``wisp'' and ``claw'' features present in the images. Wisps are caused by scattered light coming off-axis and bouncing off the top secondary mirror strut, while claws are artefacts due to scattered light coming from extremely bright stars. Minimising the impact of these two features is important not to over-subtract the background. We refer to Sect.~2.1 of \citet{zhuangActiveGalacticNuclei2024} for a complete description of the data reduction. The final mosaics released\footnote{All reduced F150W images used in this work are available at \url{https://ariel.astro.illinois.edu/cosmos_web/}.} showed an overall improved background subtraction compared to the public data release 0.2 by the COSMOS-Web team.
% \citep[see Fig. 2 of][]{zhuangActiveGalacticNuclei2024}. 
Since we used the JWST/F150W imaging data to identify mergers, we limited the COSMOS2020 catalogue to the area observed by the COSMOS-Web program and reduced by \citet{zhuangActiveGalacticNuclei2024}, as shown in Fig.~\ref{fig:COSMOS-Web}.

\subsection{X-ray, far-IR and sub-millimetre (sub-mm) data}

%The COSMOS2020 catalogue is accompanied by numerous ancillary datasets.

We included X-ray photometry from the Chandra COSMOS Legacy survey \citep{civanoChandraCosmosLegacy2016, marchesiChandraCOSMOSLegacy2016}. We selected X-ray sources from the catalogue of optical and IR counterparts presented by \citet{marchesiChandraCOSMOSLegacy2016} that have a final counterpart identification \texttt{Flag} $=1$ (secure) or $=10$ (ambiguous), and a star flag \texttt{Star} $\neq 1, 10, 100$ (which identify spectroscopically, photometrically, and visually identified stars, respectively). We then cross-matched this catalogue with our selected sources in COSMOS2020 within a radius of $1\arcsec$ of the optical coordinates. We found a total of 270 cross-matched X-ray sources.  \citet{marchesiChandraCOSMOSLegacy2016} catalogue provides only upper limits for some sources. Nevertheless, we kept these sources since the SED fitting tool employed can deal with upper limits.

%\subsection{Far-IR data: the deblended catalogue}
The following far-IR and sub-mm data are available in the COSMOS field: \textit{i)} \textit{Spitzer}/MIPS $24\mu$m data provided by the COSMOS-\textit{Spitzer} programme \citep{sandersSCOSMOSSpitzerLegacy2007}; \textit{ii)} \textit{Herschel}/PACS maps from the PACS Evolutionary Probe \citep[PEP,][]{lutzPACSEvolutionaryProbe2011} survey; \textit{iii)} \textit{Herschel}/SPIRE maps from the \textit{Herschel} Multi-tiered Extragalactic Survey \citep[HerMES][]{oliverHerschelMultitieredExtragalactic2012}. \citet{wangProbabilisticProgressiveDeblended2024} deblended these far-IR and sub-mm maps with a novel progressive and probabilistic approach. In this way, the multi-wavelength information, the full posterior, the variance, and the covariance between sources are exploited. In this paper, we used the deblended catalogue released by \citet{wangProbabilisticProgressiveDeblended2024} to add the far-IR and sub-mm data. \citet{wangProbabilisticProgressiveDeblended2024} constructed their initial prior catalogue from the COSMOS2020 catalogue. Therefore, we used the galaxy's unique IDs to cross-match the sources. We found 13\,924, 4\,392, and 1\,741 MIPS, PACS, and SPIRE counterparts, respectively. For those galaxies with a measured far-IR flux below the corresponding total noise (instrumental plus confusion noise), we set the total noise to be the flux upper limit \citep[see Table 3 in][]{wangProbabilisticProgressiveDeblended2024}.

%--------------------------------------------------------------------
\section{Methods}\label{sect:Methods}

In this section, we first describe the SED fitting method we employed, then our AGN selections, and finally, the deep learning algorithm and the mock galaxy images used to identify mergers.

\subsection{\texttt{CIGALE} SED fitting}\label{sect:CIGALE}

We estimated galaxy physical properties using the SED fitting tool Code Investigating GALaxy Emission \citep[CIGALE;][]{burgarellaStarFormationDust2005, nollAnalysisGalaxySpectral2009, boquienCIGALEPythonCode2019}. The 2022.1 version\footnote{Every \texttt{CIGALE} version is accessible at \url{https://cigale.lam.fr/}} includes AGN models and can exploit data from X-ray to radio wavelengths \citep[][]{yangXCIGALEFittingAGN2020, yangFittingAGNGalaxy2022}. The complete parameter space of the \texttt{CIGALE} configuration used is provided in Appendix~\ref{app:CIGALE}, Table~\ref{tab:cigale_parameters}. Here, we briefly describe the \texttt{CIGALE} configuration.
We employed a delayed$-\tau$ plus an optional exponential starburst star-formation history, which can model both early- and late-type galaxies, using small and large $\tau$, respectively \citep{boquienCIGALEPythonCode2019}. Moreover, including an optional exponential burst component can account for potential recent star formation. We utilised the \citet{bruzualStellarPopulationSynthesis2003} single stellar population model, with Chabrier initial mass function (IMF) and solar metallicity; the modified \citet{charlotSimpleModelAbsorption2000} as a dust attenuation law; and the \citet{draineAndromedasDust2014} models as dust emission templates. 

We used the SKIRTOR model for the AGN component, \citep{stalevski3DRadiativeTransfer2012}. SKIRTOR assumes a flared disk geometry for the dust distribution and models the dusty torus as a two-phase medium consisting of high-density clumps and a low-density medium which fills the space between the clumps in the 3D structure. We set \texttt{CIGALE} to measure the AGN fraction, $f_{\rm AGN}$, defined as the AGN contribution to the total galaxy emission, in the rest-frame wavelength range $3-30\,\mu{\rm m}$. We included the X-ray module \citet{yangXCIGALEFittingAGN2020, yangFittingAGNGalaxy2022} for modelling X-ray emission from both AGN and galaxies (due to hot gas and X-ray binaries). When only upper limits are available, these values are passed to \texttt{CIGALE} as such.

\begin{figure}
    \centering
    \includegraphics[width=.45\textwidth]{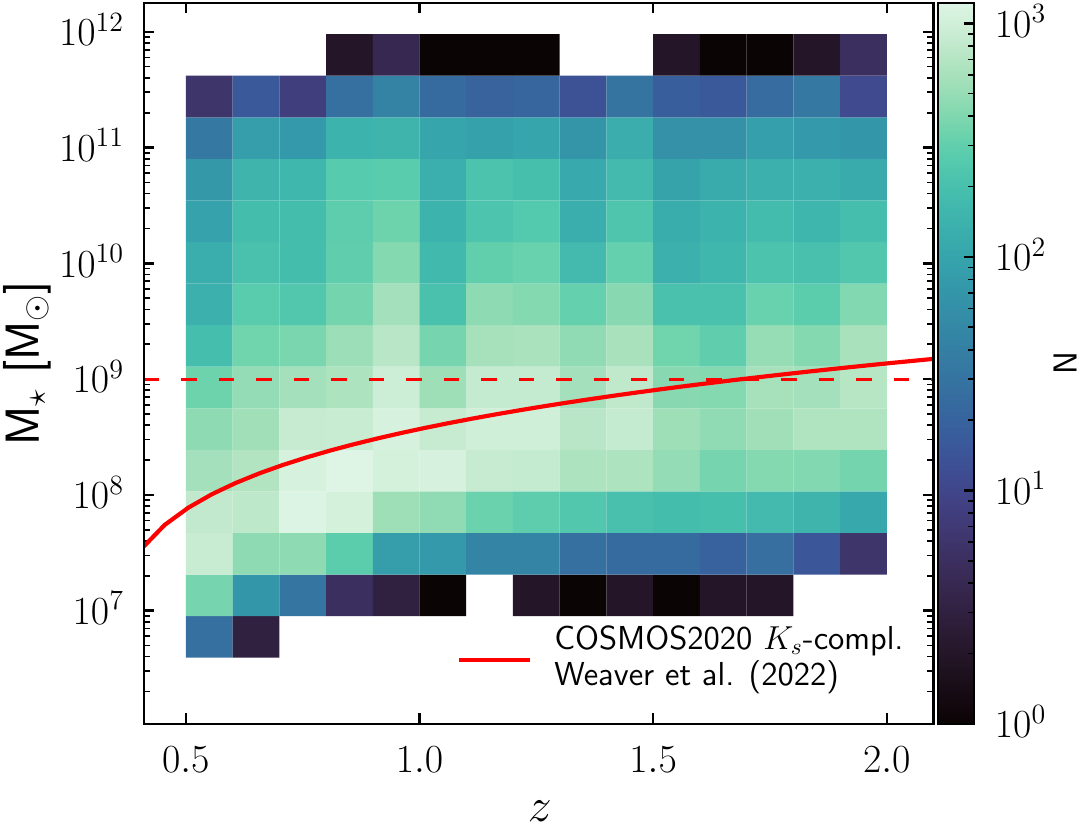}
    \caption{Stellar mass vs. redshift. The red line represents the $K_S$-based completeness for the COSMOS2020 sample \citep{weaverCOSMOS2020PanchromaticView2022}. The dashed line indicates the simulations lower mass limit (${\rm M}_{\star}=10^9{\rm M}_{\odot}$).}
    \label{fig:Mstar_z}
\end{figure}

Fig. \ref{fig:Mstar_z} shows the stellar mass distribution of the sample as a function of $z$. We selected a stellar-mass complete sample (for both quiescent and star-forming galaxies) using the $K_S$-based completeness function from \citet{weaverCOSMOS2020PanchromaticView2022} for the COSMOS2020 sample. 
%This function ensures a mass-complete sample of both quiescent and star-forming galaxies. 
As we train our DL models on galaxies more massive than $10^9\,$M$_{\odot}$, it is reasonable to set this as a lower mass limit. Therefore, we select galaxies with M$_{\star}$ higher than the maximum of $10^9\,$M$_{\odot}$ and the $K_S$-based completeness limit. 
In addition, our sample is limited to galaxies with reliable SED fits (i.e. reduced $\chi^2<5$). After applying cuts on mass and $\chi^2$, we obtained 22\,862 galaxies, excluding 2\,310 located at the edges of the JWST tiles. The final mass-complete sample consists of 20\,552 galaxies between $z=0.5$ and 2. 
The median \texttt{CIGALE} error is $\sigma (\log_{10}{\rm M}_{\star}) = 0.1$ for stellar mass , $\sigma (\log_{10}{\rm SFR}) = 0.2$ for SFR, and $\sigma(f_{\rm AGN}) = 0.1$ for AGN fraction. %The median error output by \texttt{CIGALE} for the AGN fraction is $\sigma(f_{\rm AGN}) = 0.1$. 
We further assessed the reliability of galaxy property measurements in Appendix~\ref{app:CIGALE}, with example best-fit SEDs. 

Additionally, we ran \texttt{CIGALE} without SKIRTOR and X-ray modules, keeping all other parameters as in Table~\ref{tab:cigale_parameters}. This test was meant to check whether galaxies could be fit equivalently well without an AGN component. The output results helped in identifying a purer and more reliable sample of SED AGN, as described in the next subsection.

\subsection{AGN selections}\label{sect:agn_sel}

\begin{figure}
    \centering
    \includegraphics[width=.49\textwidth]{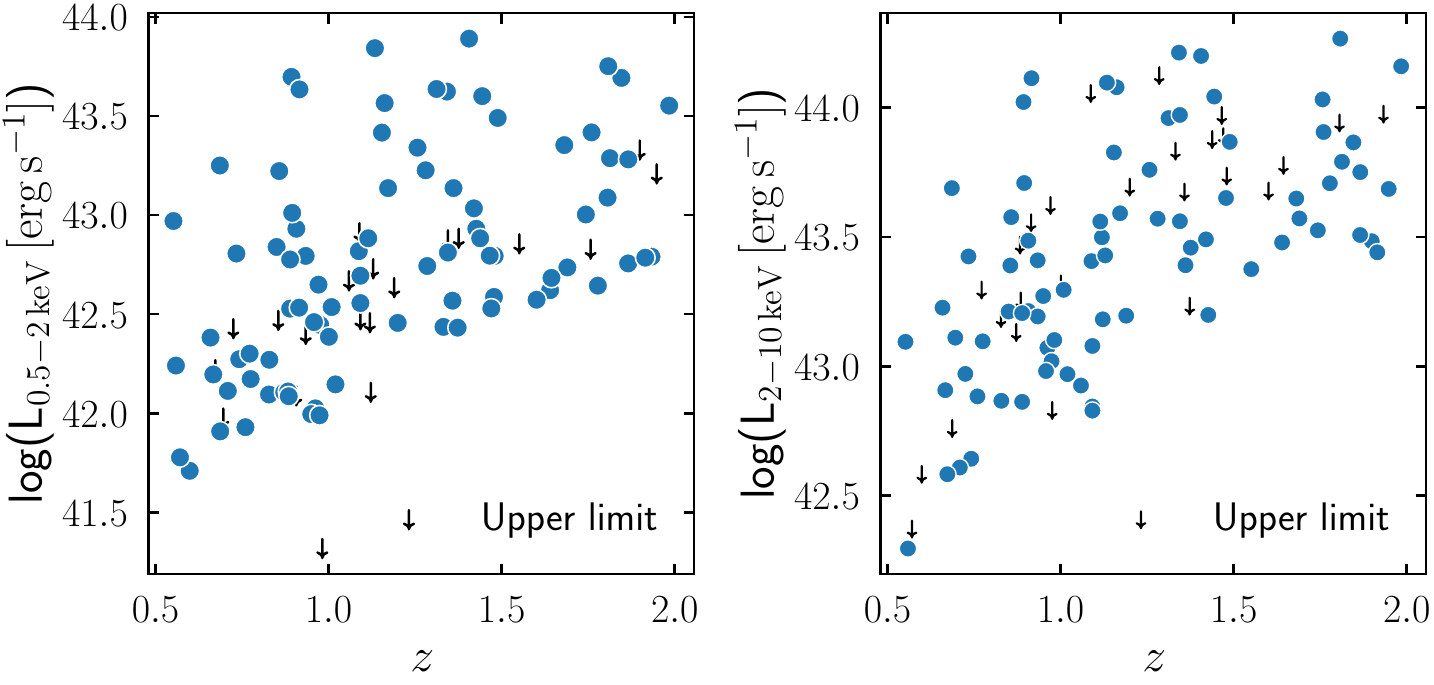}
    \caption{Rest-frame soft (left) and hard (right) X-ray luminosities vs. redshift for the selected X-ray AGN. Black arrows are upper limits. 
    } 
    \label{fig:Lxray}
\end{figure}

\begin{figure}
    \centering
    \includegraphics[width=.49\textwidth]{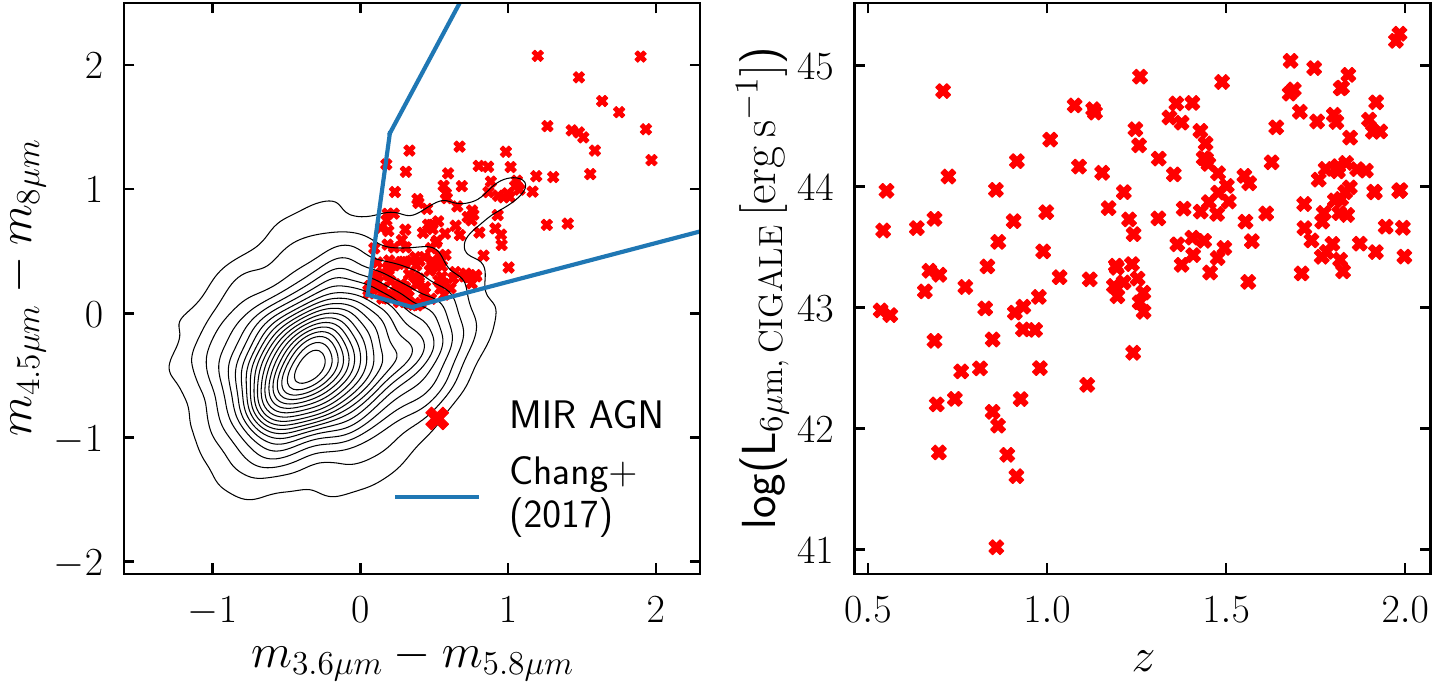}
    \caption{\textit{Left}: IRAC colour--colour space. Red crosses are the selected MIR AGN. Contours show the population satisfying the $24\,\mu$m flux and S/N requirements. The blue box indicates the \citet{changInfraredSelectionObscured2017} selection. 
    \textit{Right}: Rest-frame $6\,\mu{\rm m}$ luminosity of the AGN component vs. redshift for the selected MIR AGN. 
    }
    \label{fig:MIR_AGN}
\end{figure}

We selected AGN in three different ways. 
% X-ray AGN selection. 
First, following \citet{marchesiChandraCOSMOSLegacy2016}, we selected 104 secure X-ray AGN by requiring \texttt{DET\_ML} (the maximum likelihood detection) $>10.8$ in the hard or soft band. In about half of the cases (49/104) \texttt{DET\_ML} is $>10.8$ in both bands. 
%For the other half of the sources, one of the two bands has only an upper limit flux estimate. 
In Fig.~\ref{fig:Lxray}, we show the X-ray luminosity as a function of redshift, in both soft and hard bands, for the selected X-ray AGN. No sources in the hard band have luminosities ${\rm L}_X<10^{42}\, {\rm erg\,s^{-1}}$, which is the threshold conventionally used to identify clear AGN from galaxies \citep[][]{rosarioMeanStarFormation2012}. In the soft band, roughly $7\%$ of the sources have $10^{41}<{\rm L}_X<10^{42}\,{\rm erg\,s^{-1}}$. However, excluding these sources does not affect our results, so we keep them in the sample.

% MIR AGN
Second, we identify AGN by their MIR emission in galaxies with \emph{Spitzer/}MIPS $24\,\mu{\rm m}$ flux $F_{24\,\mu{\rm m}}>20\, \mu{\rm Jy}$, which is the $1\sigma$ total noise (instrument plus confusion noise). Then, we required $S/N >5$ in each IRAC channel and applied the colour--colour criteria in \citet{changInfraredSelectionObscured2017}\footnote{Equation~3 presented in \citet{changInfraredSelectionObscured2017} contains an error. After discussion with the authors, we report the correct version here.}:
\begin{align}
    y &< 2.22 x +1.01, \\
    y &< 8.67 x -0.28, \\
    y &> -0.33 x +0.17, \\
    y &> 0.31 x -0.06,
\end{align}
where $x=m_{3.6\,\mu {\rm m}} - m_{5.8 \,\mu {\rm m}}$ and $y=m_{4.5\,\mu {\rm m}} - m_{8\, \mu {\rm m}}$. Magnitudes are in the AB system. This colour--colour selection is displayed in Fig.~\ref{fig:MIR_AGN}, left panel. In total, we found 159 MIR AGN. We show their rest-frame $6\,\mu$m luminosity in the right panel of Fig.~\ref{fig:MIR_AGN}.

% SED AGN
Third, we defined SED AGN as those galaxies where the inclusion of an AGN component significantly improves the fit to the observed photometry. To assess the necessity of the AGN component, we compared the \texttt{CIGALE} runs which include the AGN module with the run excluding the AGN module. We classified a source as an SED AGN only if the relative improvement in the reduced $\chi^2$ exceeded 10\%, or the fit without AGN component failed. Furthermore, to ensure that the AGN identification is physically constrained by the hot dust emission rather than fitting degeneracies in the rest-frame near-infrared, we imposed a strict MIR photometric requirement: a detection with $S/N>5$ in all four \textit{Spitzer}/IRAC channels (3.6, 4.5, 5.8, and 8.0 $\mu$m). Finally, we required an AGN fraction $f_{\rm AGN}>0.2$. This threshold exceeds the typical \texttt{CIGALE} median uncertainty of 0.1, minimising misclassifications, while the combination of the $\chi^2$ evaluation and MIR coverage criteria ensures the presence of a robust MIR excess due to the presence of an AGN. 

\begin{figure}
    \centering
    \includegraphics[width=.3\textwidth]{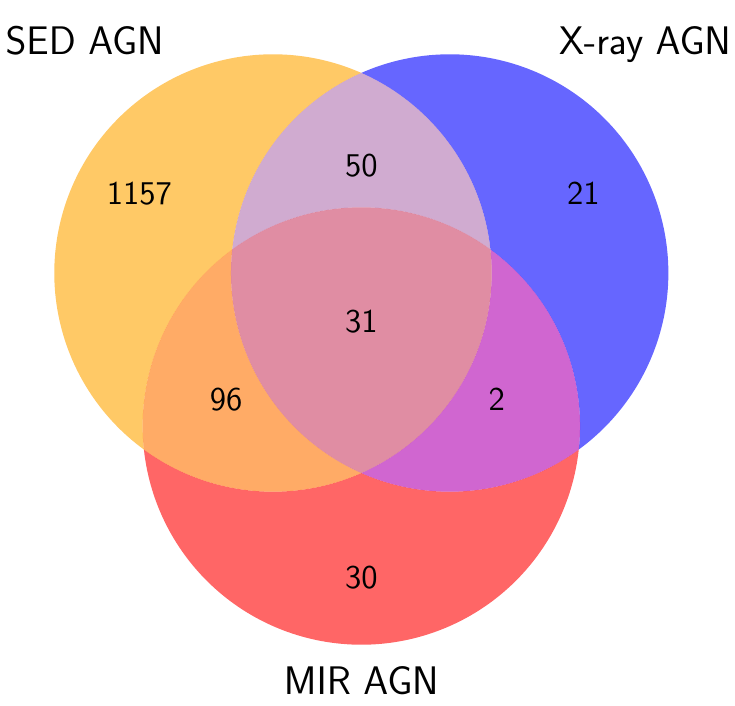}
    \caption{
    Venn diagram of the three AGN types. 
    }
    \label{fig:venn}
\end{figure}

In summary, SED AGN respect the following three criteria:
\begin{align}
\label{eq:SED_AGN}
    & \frac{\chi^2_{\rm red.,\,noAGN}-\chi^2_{\rm red.,\, AGN}}{\chi^2_{\rm red.,\, AGN}} >10\%\,{\rm {\texttt OR} \, failed\, fit\, without\, AGN} ,\\
    & S/N_{\rm IRAC\, CH1,\, CH2,\, CH3,\, CH4}>5\, , \\
    & f_{\rm AGN} >0.2\,.
\end{align}
We found 1334 SED AGN in our sample. Figure~\ref{fig:venn} summarises the numbers and overlaps between the categories in our sample. Although our SED AGN incorporate most of the MIR and X-ray AGN ($\approx 80\%$), these two diagnostics still identify unique AGN that are missed by the SED definition.

\begin{figure}
    \centering
    \includegraphics[width=.49\textwidth]{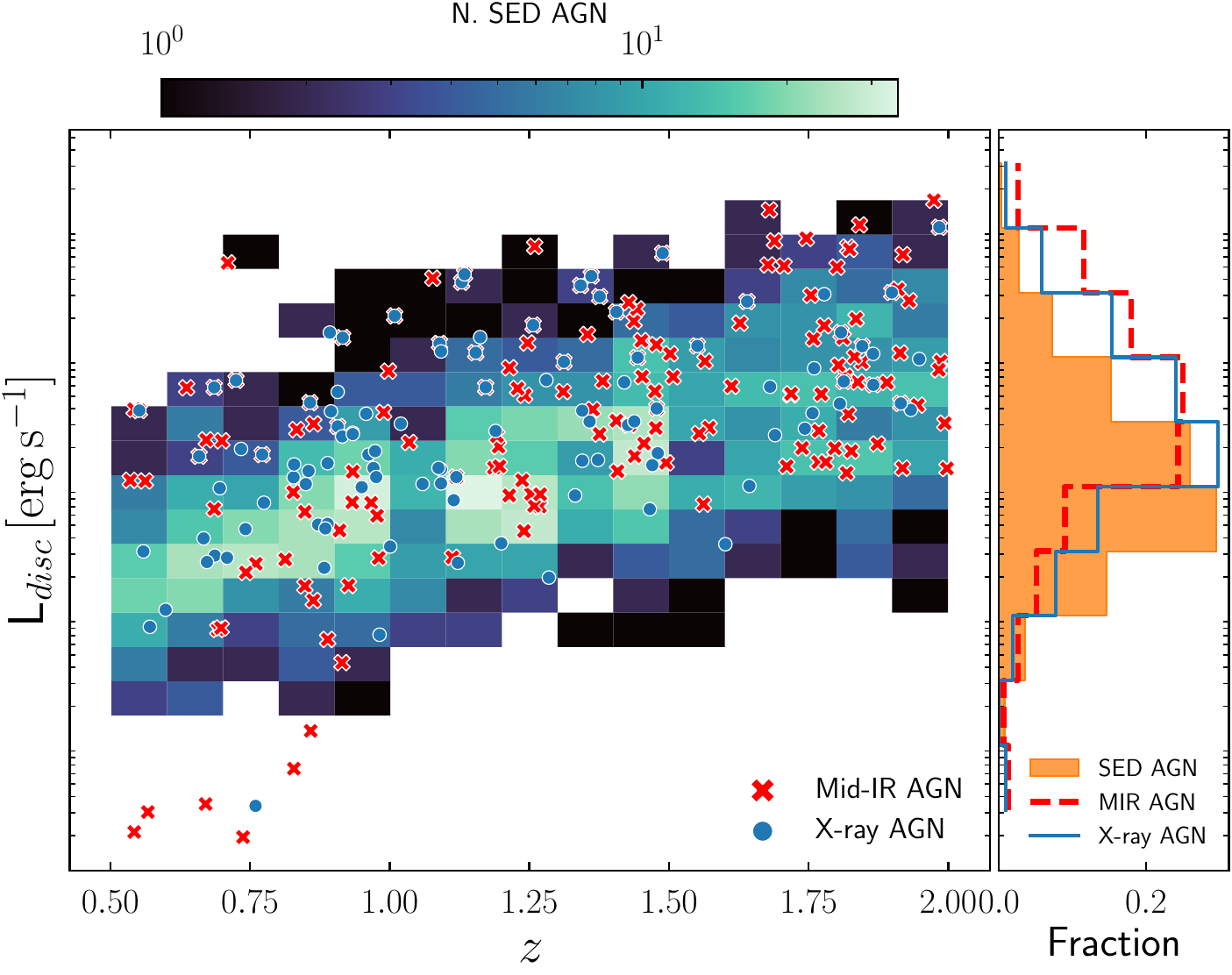}
    \caption{
    AGN disc accretion luminosity as a function of redshift, for MIR (red crosses), X-ray (blue circles) and SED AGN (2D histogram). On the right margin, we display the L$_{disc}$ histogram distributions for the three AGN selections. 
    }
    \label{fig:SED_AGN}
\end{figure}

\begin{figure}
    \centering
    \includegraphics[width=0.44\textwidth]{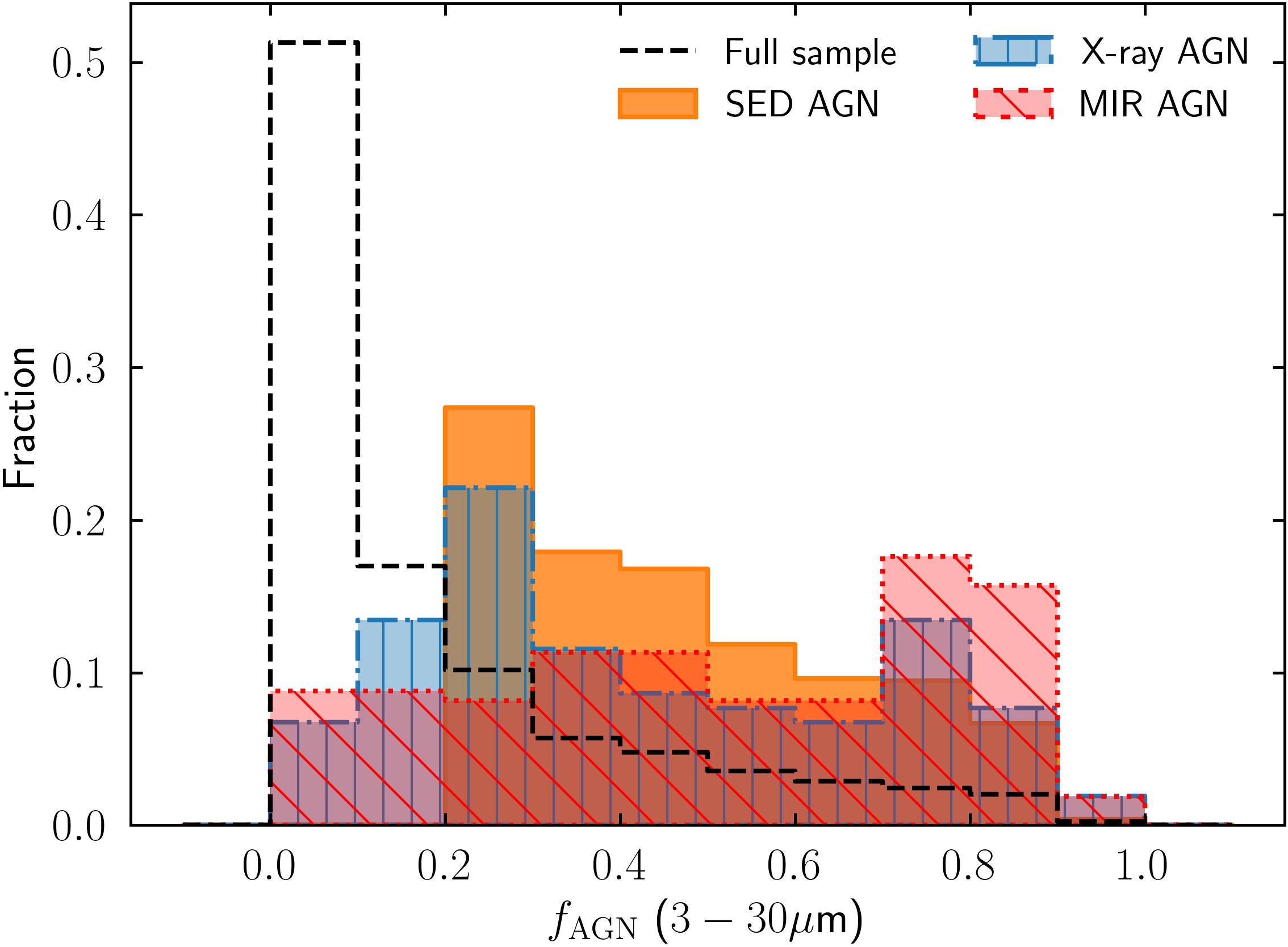}
    \caption{
    Normalised AGN fraction distribution for the three AGN classes (SED: orange bars; X-ray: blue bars with vertical stripes; MIR: red bars with diagonal stripes) and the entire galaxy sample (dashed black line). }
    \label{fig:fagn_dist_AGN}
\end{figure}

We show the AGN luminosity distribution for the three classes in Fig.~\ref{fig:SED_AGN}. As a proxy for the bolometric luminosity, we used L$_{disc}$, the ``accretion power'' parameter from \texttt{CIGALE}, which is equivalent to the angle-averaged AGN bolometric luminosity \citep{yangLinkingBlackHole2018}. In general, MIR and X-ray AGN exhibit intermediate-to-high L$_{disc}$, with the MIR AGN constituting the brightest sample on average. Conversely, the SED AGN selection extends to a larger population of faint source, particularly at $z>1.2$, although the majority of this sample still shows intermediate luminosities.
We show the $f_{\rm AGN}$ distributions for the three AGN selections and the entire galaxy sample in Fig.~\ref{fig:fagn_dist_AGN}. The galaxy population peaks at low $f_{\rm AGN}$ values (mostly $<0.20$) and monotonically decreases with increasing $f_{\rm AGN}$. The SED AGN follow this trend, but starting at $f_{\rm AGN}\geq 0.20$. The MIR and X-ray AGN show different distributions. The MIR AGN distribution is roughly flat, with a slight skew towards high $f_{\rm AGN}$ values. This is expected as we measure $f_{\rm AGN}$ in the rest-frame $3-30\,\mu$m range. The X-ray distribution is also broadly uniform, with a peak at $f_{\rm AGN}\approx0.2-0.3$.

\subsection{Merger identification}\label{sect:CNN}

We used convolutional neural networks (CNNs) trained on realistic mock images of simulated galaxies as our merger classifier. 

\subsubsection{Mock galaxy images}\label{sect:simul}

The IllustrisTNG cosmological hydrodynamical simulation consists of three different volumes varying in physical size and mass resolution \citep{marinacciFirstResultsIllustrisTNG2018, naimanFirstResultsIllustrisTNG2018, nelsonFirstResultsIllustrisTNG2018, pillepichFirstResultsIllustrisTNG2018, springelFirstResultsIllustrisTNG2018}. We used the highest resolution version of the TNG-100 box (hereafter referred to as TNG), whose side corresponds to $\approx 110.7\,{\rm Mpc}$ and has a baryonic matter resolution of $1.4\times10^6\,{\rm M_{\odot}}$. This mass resolution allowed us to select galaxies down to M$_{\star}=10^9\,{\rm M_{\odot}}$. Galaxies were selected in the range $z=0.5$--$2$ (corresponding to snapshot numbers $67-33$). For each galaxy identified through the \texttt{Subfind} algorithm \citep{springelPopulatingClusterGalaxies2001}, TNG provides a complete merger history  \citep{rodriguez-gomezMergerRateGalaxies2015}.

% Horizon-AGN
Horizon-AGN is a cosmological hydrodynamical simulation of a 100 Mpc$^3\,h^{-1}$ comoving volume. It has a stellar mass particle resolution of $2\times10^6\,{\rm M_{\odot}}$ \citep{duboisDancingDarkGalactic2014},
% {duboisHORIZONAGNSimulationMorphological2016}, 
comparable to that of TNG-100. We identified galaxies with the \texttt{AdaptaHOP} algorithm, updated to construct merger trees \citep{tweedBuildingMergerTrees2009}. In this case, galaxies were also selected in the  $z=0.5$--$2$ range and with M$_{\star}>10^9\,{\rm M_{\odot}}$. 
For both simulations, we observed each object from three different projections. 

We defined mergers as galaxies that had a major merger (with a stellar mass ratio $\leq 4$) in the last 300 Myr and/or will coalesce in the next 800 Myr. Galaxies not meeting these criteria were labelled as non-mergers. We found 45\,850 and 42\,630 mergers in TNG and Horizon-AGN, respectively. We selected the same number of non-mergers in both simulations, roughly matching the stellar mass and redshift distributions of mergers. We created mock F150W observations of simulated galaxies from three different points of view. To obtain reliable classifiers, it is crucial to include observational effects \citep[][]{huertas-companyHubbleSequence02019, rodriguez-gomezOpticalMorphologiesGalaxies2019}. 
We generated synthetic observations following \citet{margalef-bentabolGalaxyMergerChallenge2024}:
\begin{enumerate}
    \renewcommand{\labelenumi}{\it \roman{enumi})}
    
    \item We created galaxy thumbnails with a physical size of 50 kpc $\times$ 50 kpc and the same pixel resolution as the real NIRCam images ($0.03\arcsec/$pixel). Each stellar particle contributes to the galaxy's SED, determined by its mass, age, and metallicity. We derived these SEDs from the stellar population synthesis models of \citet{bruzualStellarPopulationSynthesis2003}. The integrated SED was passed through the F150W filter to generate a smoothed 2D projected map \citep{rodriguez-gomezOpticalMorphologiesGalaxies2019}. 
    
    \item We convolved each image with the filter point spread function (PSF), randomly choosing one from the 80 PSF models derived by \citet{zhuangActiveGalacticNuclei2024}. We then added Poisson noise to the convolved images as shot noise. 
    
    \item To make observations as realistic as possible, each mock observation was injected into real F150W sky cutouts. We created F150W cutouts that do not contain any bright source at the centre and without artifacts but still allow for possible background galaxies and faint sources. We generated random coordinates within the covered area and ensured that there were no $z<3$ sources in COSMOS2020 within a radius of $6\farcs5$. This radius was derived from the estimated source density of COSMOS2020. The generated coordinates were used as the centres of our cutouts, which were $320$ pixels across (corresponding to $\sim 60\,\rm{kpc}$ at $z=0.5$). To ensure that no stellar spikes are present in the cutouts, we ran Kendall's $\tau$ test \citep{kendallNEWMEASURERANK1938} along both image axes. If a strong correlation among pixels was found, that is, a $p$-value $<0.001$, the cutout was rejected.\footnote{The code used for this step is publicly available at \url{https://github.com/Antonio-LM/Imaging-pipeline}} For each mock galaxy, we randomly picked a sky cutout and injected the galaxy into its centre. 
\end{enumerate}
Although we previously explored incorporating dust via radiative transfer, we found the computational cost to be extremely high, while the impact on the resulting morphologies---and on classification performance---was minimal \citep{bottrellDeepLearningPredictions2019, rodriguez-gomezOpticalMorphologiesGalaxies2019, wangConsistentFrameworkComparing2020}. Therefore, we opted to use dust-free simulations in this work.

To account for potentially bright AGN, we included a central point source in $20\%$ of the simulated galaxies (randomly selected), using the the PSF models derived by \citet{zhuangActiveGalacticNuclei2024}. 
As performed by \citet{margalef-bentabolAGNHostGalaxy2024}, the PSF contribution fraction (with values drawn uniformly between 0 and 1) was defined in relation to the host galaxy,
\begin{equation}
    f_{\rm{PSF}} = \frac{F_{\rm{PSF}}}{F_{\rm{host}} + F_{\rm{PSF}}} \;,
\end{equation}
where $F_{\rm{PSF}}$ and $F_{\rm{host}}$ are the fluxes within a $0\farcs5$ aperture of the central source and the host galaxy, respectively.  

Our CNN takes as input images of the same size. Thus, we resized all images to a common size of $256$ pixels across, corresponding to $\sim 50\,\rm{kpc}$ at $z=0.5$. Following \citet{bottrellDeepLearningPredictions2019}, all images were hyperbolic arcsin-scaled in the range $0$--$1$, and the contrast of the central target was maximised as follows:
\begin{enumerate}
  \renewcommand{\labelenumi}{\it \roman{enumi})}
  \item We took the hyperbolic arcsin of the sky-subtracted images. Values below $-7$ were converted to NaNs. 
  \item We computed the median of each image, $a_{\rm min}$, and the 99th percentile, $a_{\rm max}$, considering a central box of side 80 pixels. 
  \item All values $<a_{\rm min}$ were set to $a_{\rm min}$, including the NaNs. Values $>a_{\rm max}$ were set to $a_{\rm max}$. Then, the clipped images were normalised by subtracting $a_{\rm min}$ and dividing by $a_{\rm max}-a_{\rm min}$.
\end{enumerate}

\subsubsection{CNN training}\label{sect:cnn}

We developed a CNN to classify galaxies into mergers and non-mergers, utilising the Keras framework for the TensorFlow platform \citep{cholletKerasDeepLearning2023, abadiTensorFlowLargeScaleMachine2016} for the architecture implementation.
% CNNs consist of multiple layers of neurons which perform various operations. The output map of each layer is passed to the next layer. Lower (initial) layers convolve the input image with a predefined number of filters, made of sets of neurons with trainable weights and biases. Higher layers are typically 1D fully-connected layers, where all neurons are connected with the neurons of the previous layer. The final CNN output is a score for each input image. The predicted score is then used to make the classification. Training a CNN requires a large amount of labelled images to adjust the weights to match the given classification. 
 % 
% The architecture developed in this work is presented in Table~\ref{tab:CNN}, for which we utilised the Keras framework for the TensorFlow platform \citep{cholletKerasDeepLearning2023, abadiTensorFlowLargeScaleMachine2016}. 
The CNN consists of four convolutional layers and three fully-connected layers. The output map of each layer is passed to the next layer. For all layers, we adopted a Rectified Linear Unit as an activation function. A stride of one pixel was used for the convolutional layers. We introduced dropout layers after each processing layer to prevent over-fitting. These dropout layers randomly set input units to 0 at a rate specified by the user. To further prevent over-fitting, early stopping in the training phase was used. The architecture and specific hyper-parameters are reported in Appendix~\ref{app:CNN}, Table~\ref{tab:CNN}. 

We trained an identical CNN using mock galaxy observations from both the TNG and Horizon-AGN simulations. We designate the CNN trained on TNG images as TNG-CNN and the one trained on the Horizon-AGN dataset as HA-CNN. The CNN output is a score for each input image, in turn, used to make the classification. We split each mock galaxy sample into train, validation, and test sets, using a 80--10--10 split. We initially evaluated each model performance on its associated test set using common metrics such as ``precision'', ``recall'', and ``F1-score'' calculated for the merger class. TNG-CNN has a precision of 0.74, a recall of 0.61, and an F1-score of 0.67 on the TNG test set. Overall, HA-CNN shows better performance with a precision of 0.76, recall of 0.69, and F1-score of 0.72.

% We evaluated the model performance on the test set from
% the TNG simulations, using common metrics such as “precision”,
% “recall”, and “F1-score”. Precision measures how often
% the model correctly predicts a given class, while recall focuses on
% how complete the model is at finding objects in a given class. In
% other words, precision is the number of objects correctly recovered
% for a class divided by the total number of objects predicted
% in that class. Recall is the number of objects correctly recovered
% for a class divided by the total number of objects in that class.
% F1-score is the harmonic mean of precision and recall. All metrics
% are calculated on a balanced sample (50% mergers and 50%
% non-mergers).

\subsubsection{Predicting on JWST/F150W images}

To distinguish morphological features reliably, we restricted the sample to those with $S/N> 10$. We calculated the $S/N$ by performing aperture photometry with a circular aperture centred on each galaxy, with a radius of $2.5\,{\rm kpc}$ at the galaxy $z$. The background noise is calculated using 500 blank cutouts. In each blank cutout, we applied sigma clipping ($\sigma = 3.0$) and calculated the root mean square. Then, we computed the median background noise of the 500 cutouts. We found $13\,789$ galaxies with $S/N>10$ in the range $z\in [0.5; 2.0]$. This is the sample of galaxies on which we focus in this paper. 

\begin{table}[h]
    \centering
    \caption{Overall CNN performance on the visual test set.}
    \label{tab:performance}
    \small
    \begin{tabular}{lccc}
    \hline\hline \\[-7pt]
    Metric & TNG-CNN & HA-CNN & Comb-CNN \\
    \hline\\[-7pt]
    Accuracy & 0.765 & 0.758 & 0.885 \\
    Precision & 0.805 & 0.820 & 0.876\\
    Recall & 0.701 & 0.664 & 0.897 \\
    F1 score & 0.750 & 0.734 & 0.886\\
    \hline
    \end{tabular}
    \tablefoot{Precision, Recall, and F1 score refer to the merger class.}
\end{table}

We evaluated the performance of the models using a test set of real JWST/F150W images. For this task, three of us (ALM, LW, BMB) visually inspected 2000 galaxies, randomly sampled in the range $0.5\leq z \leq 2$. A galaxy was labelled as a merger or non-merger only if all three classifiers agreed on the class; otherwise, the galaxy was rejected. The identification of mergers was based on the presence of a companion of comparable size, morphological disturbances such as tidal tails or bridges, and highly irregular structures indicative of recent or ongoing interactions. 

The final visually inspected sample contained 891 non-mergers and 378 mergers and was used to measure the performance of the CNNs on the real observations. This test set was balanced to obtain an even number of mergers and non-mergers, randomly sampling 378 non-mergers among the 891 available. This step was repeated multiple times, but the final results did not change. For both models, we searched for the best threshold to divide mergers and non-mergers, defined as the threshold which maximises the F1 score, maintaining a precision $>0.80$ for the merger class. The results are reported in Table~\ref{tab:performance}. Overall, the two models show similar good performance on the visually inspected test set, with precision of $\simeq 0.8$ and recall of $\simeq 0.7$ for the merger class.

\subsubsection{The combined CNN classifier}

\begin{figure}
    \centering
    \includegraphics[width=0.45\textwidth]{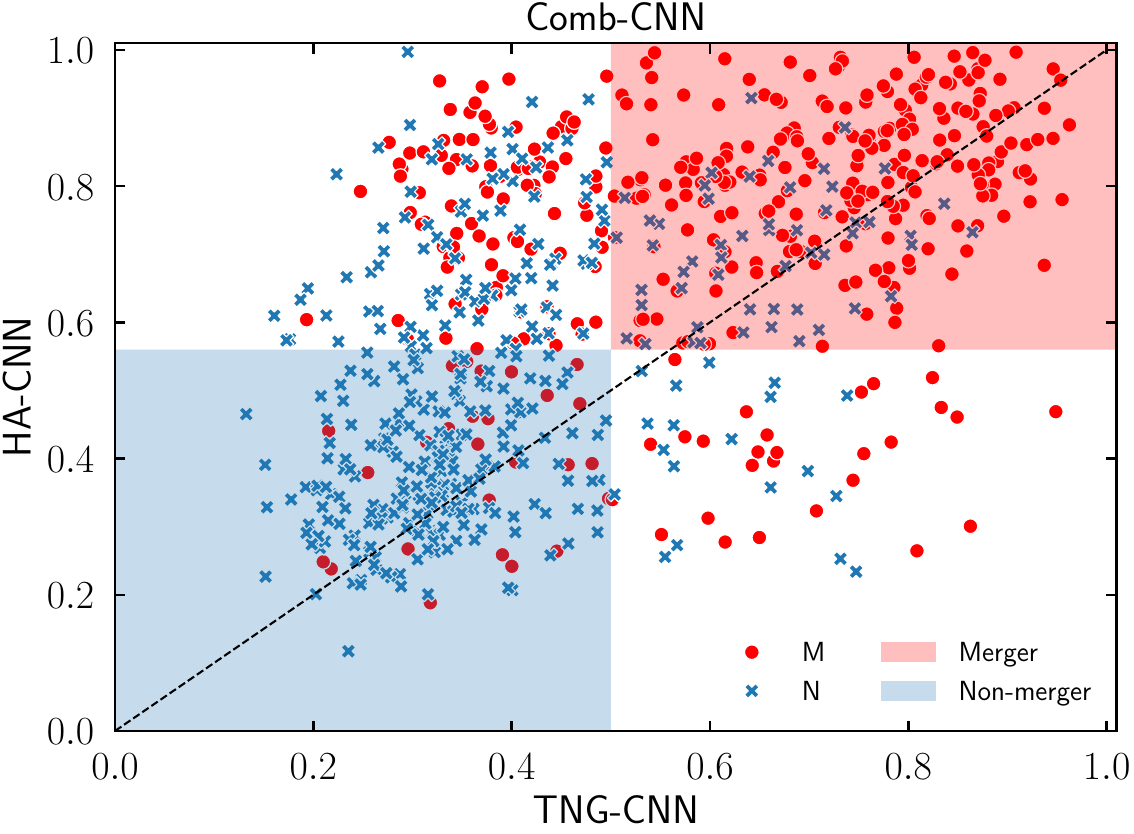}
    \caption{The Comb-CNN classification for the visually inspected test set (mergers: red circles; non-mergers: blue crosses). 
    The scores predicted by the TNG-CNN and the HA-CNN are reported on the $x$- and $y$-axes, respectively. The shaded red/blue area defines the region of galaxies classified as mergers/non-mergers by the Comb-CNN. 
    Sources in the white areas are labelled as unclassified. 
    }
    \label{fig:Comb-CNN}
\end{figure}

\begin{figure}[h]
    \centering
    \includegraphics[width=0.34\textwidth]{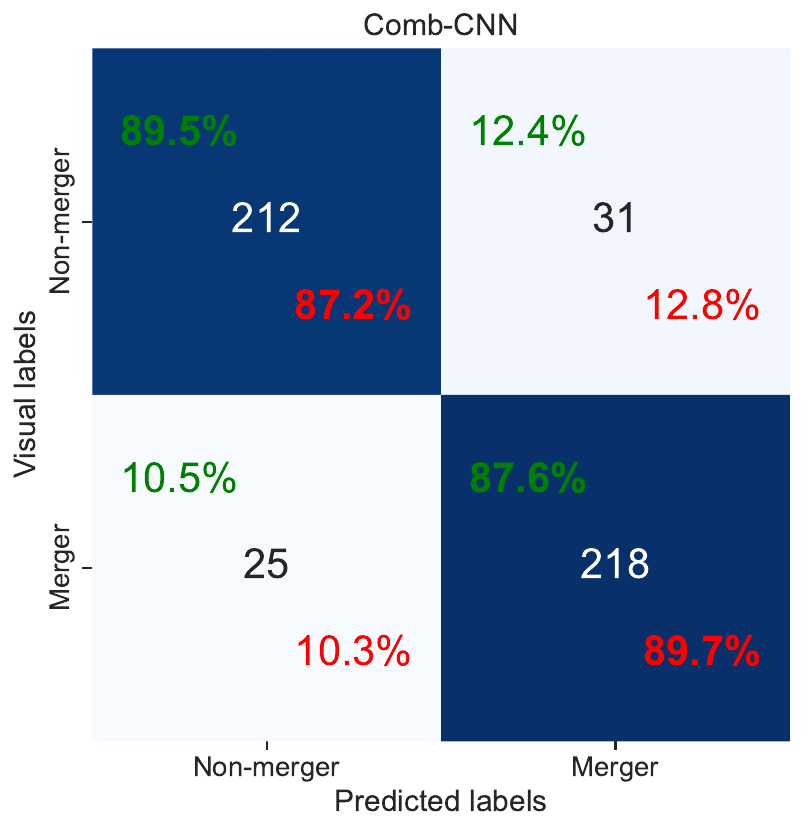}
    \caption{The Comb-CNN confusion matrix for the visually inspected test set. Each cell shows the number of objects at its centre, the fraction of objects normalised by column (green percentages) in the top left corner, and the fraction of objects normalised by row (red percentages) in the lower right corner. On the main diagonal, these percentages correspond to precision and recall, respectively.}
    \label{fig:CM}
\end{figure}

We followed \citetalias{lamarcaDustPowerUnravelling2024a} to have even higher classification precision by combining the TNG-CNN and the HA-CNN into a single classifier (hereafter Comb-CNN). Comb-CNN works as a logic \texttt{AND} operator, as shown in Fig.~\ref{fig:Comb-CNN}: if both CNNs agreed on the class (merger or non-merger), then the image was classified as such, otherwise, it was labelled as ``unclassified''. We searched for the best thresholds using a similar approach as for the individual models. We looked for the combination of two thresholds ($T_{\rm TNG}$ and $T_{\rm HA}$) that gives the best F1 score, with the constraint of a merger class precision $>0.8$. Additionally, as Comb-CNN excludes some galaxies, we ensured at every step that the test set contained at least 250 visually classified mergers and 250 non-mergers. For each threshold combination, the test set was re-balanced 20 times, and the metrics were calculated as the median value of the 20 drawings. Each threshold was varied in the range of $0.3$--$1.0$, with a step of 0.01. As the best combination of thresholds, we found $T_{\rm TNG}=0.50$ and $T_{\rm HA} = 0.56$. The performance of Comb-CNN with these thresholds is reported in Table~\ref{tab:performance}. Comb-CNN performs better than the individual models, having a merger-class precision of $0.88$ and a recall of $0.90$. We show the confusion matrix of Comb-CNN with the best threshold combination in Fig.~\ref{fig:CM}.

We used Comb-CNN and the best thresholds to classify all selected galaxies. Out of the initial $13\,789$ galaxies, $8\,285$ have been classified as mergers ($2\,276$) or as non-mergers ($6\,009$), while $5\,504$ have been labelled as unclassified. Hereafter, we focus on the sample of classified galaxies. In Appendix~\ref{app:images}, we show some examples of galaxies identified as mergers and as non-mergers by our classifier. 

\begin{figure*}
\centering
\includegraphics[width=.84\textwidth]{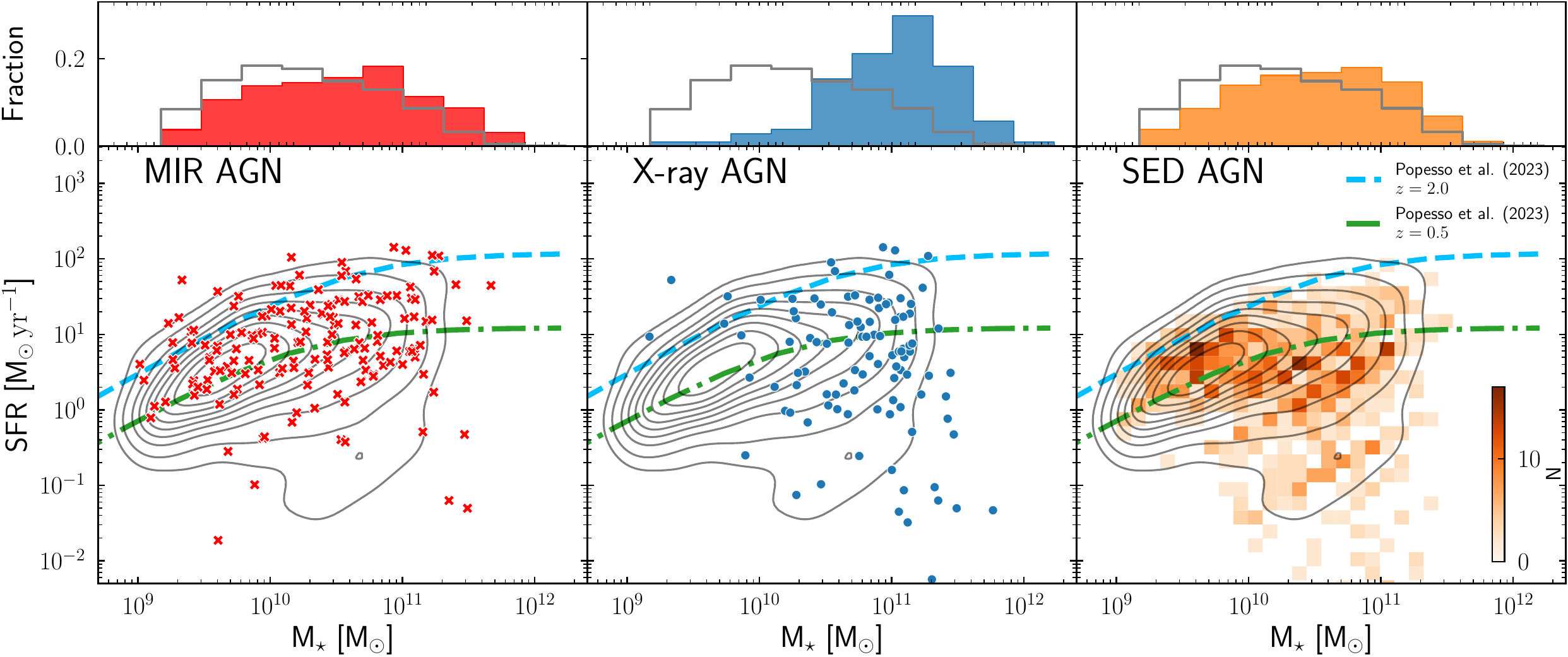}
\caption{
    SFR vs. M$_{\star}$ for the three AGN types. We show the SED AGN as a 2-D distribution. In each panel, the contours (from 10\% to 90\%, with intervals of 10\%) represent the non-AGN distribution. 
    The dashed blue and the dash-dotted green lines indicate the MS at $z=2$ and $z=0.5$ from \citet{popessoMainSequenceStarforming2023}, respectively. On top of each panel is the M$_{\star}$ distribution of the AGN and the corresponding non-AGN (grey). }
    \label{fig:AGN_MS}
\end{figure*}

%--------------------------------------------------------------------
\section{Results}\label{sect:Results}

In this section, we present the properties of the AGN host galaxies, demonstrating the importance of constructing proper control samples. Then, we investigate the merger--AGN relation adopting a simple binary AGN/non-AGN classification and a continuous approach, exploring the relative and absolute AGN power.

\subsection{The host galaxies of AGN and non-AGN}

We defined non-AGN galaxies as those not identified as X-ray or MIR AGN and with $f_{\rm AGN}<10\%$. 
%We adopted this more stringent criterion to avoid potential contamination in the non-AGN sample.
In Fig.~\ref{fig:AGN_MS}, we show the SFR--$\rm{M}_{\star}$ plane and the M$_{\star}$ distributions for the three AGN types and the non-AGN. The X-ray AGN inhabit intermediate and high-mass galaxies, with a median value of $10^{10.9}\,{\rm M_{\odot}}$. Many X-ray AGN are below the MS, appearing to be transitioning towards quiescence (the ``green valley''). MIR AGN tend to be hosted by intermediate-mass galaxies (with a median M$_{\star}$ of $10^{10.3}\,{\rm M_{\odot}}$) on the MS, with the lowest fraction of quenched galaxies. 
%This difference in SFRs might simply reflect the difference in the M$_{\star}$ distributions and the evolution of the quenched fraction with mass. 
Host galaxies of SED AGN exhibit a M$_*$ distribution comparable to that of MIR AGN (median M$_{\star}$ of $10^{10.3}\,{\rm M_{\odot}}$), and generally follow the MS, while showing a secondary overdensity in the quenched/red-sequence region. In comparison, non-AGN have larger fractions of low-mass galaxies (with a median value $\sim 10^{9.9}\,{\rm M_{\odot}}$), and display both a star-forming MS and a passive red sequence.
%A very small fraction of M$_{\star}>10^{11}\,{\rm M_{\odot}}$ galaxies do not contain any AGN. 
In Appendix~\ref{app:SFMS}, we further investigate the SFR--M$_{\star}$ relation as a function of the AGN luminosity and for exclusive types of AGN.

Our stellar mass distributions of the AGN host galaxies agree with previous results. \citet{bongiornoAccretingSupermassiveBlack2012} showed that the mass distribution of the previous generation of X-ray AGN in COSMOS peaks at $10^{10.9}\,{\rm M_{\odot}}$. \citet{mountrichasComparisonStarFormation2022} selected X-ray AGN using the eROSITA Final Equatorial-Depth Survey (eFEDS) and found a median host M$_{\star}$ of $10^{11}\, {\rm M_{\odot}}$. Selecting X-ray AGN from the XMM-XXL survey, \citet{mountrichasGalaxyPropertiesType2021} found an average host M$_{\star} \simeq 10^{10.9}\,{\rm M_{\odot}}$ for type I AGN, and M$_{\star} \simeq 10^{10.6}\,{\rm M_{\odot}}$ for type II AGN. Likewise, \citet{rosarioXRaySelectedAGN2013} found that the peak in the AGN host galaxy mass distribution is at $10^{10.5-10.7}\, {\rm M_{\odot}}$. \citet{bornanciniPropertiesIRselectedActive2022} and \citet{azadiMOSDEFSurveyAGN2017} observed a host mass distribution for MIR AGN comparable to the one we present, with a median value at $10^{10.5}\, {\rm M_{\odot}}$. In a recent study of optically selected Type II AGN, \citet{vietriTypeIIAGNhost2022} found that the stellar masses of AGN hosts have a median value of $10^{9.5}\, {\rm M_{\odot}}$, slightly lower than what we observe for the MIR and SED AGN.

The location of the X-ray AGN in the SFR--$\rm{M}_{\star}$ plane agrees with what \citet{cristelloInvestigatingStarFormation2024a} observed. Similarly, \citet{bongiornoAccretingSupermassiveBlack2012} and \citet{azadiPRIMUSRelationshipStar2015} found that X-ray AGN often inhabit massive, red galaxies. \citet{mullaneyALMAHerschelReveal2015} and \citet{silvermanEvolutionAGNHost2008} found that the host galaxies of X-ray AGN are on average below the MS. Previous studies also uncovered a SFR dependence on X-ray luminosity L$_X$, with SFRs lower or comparable to those of star-forming galaxies at L$_X<10^{44}\,{\rm erg\,s^{-1}}$, and slightly enhanced SFRs at higher L$_X$ \citep{rosarioMeanStarFormation2012, santiniEnhancedStarFormation2012, mountrichasGalaxyPropertiesType2021, mountrichasComparisonStarFormation2022}. These results seem to be in contrast with our Fig.~\ref{fig:AGN_MS}. However, \citet{mountrichasGalaxyPropertiesType2021, mountrichasComparisonStarFormation2022} excluded quiescent systems. \citet{rosarioMeanStarFormation2012} derived mean SFRs through stacking, which can be biased towards the star-forming population compared to the median distribution, while \citet{santiniEnhancedStarFormation2012} considered only galaxies detected in the far-IR. 
\citet{ellisonStarFormationRates2016} found that MIR AGN have, on average, enhanced SFRs compared to non-active galaxies and are frequently hosted by massive star-forming galaxies, in agreement with our results. \citet{azadiMOSDEFSurveyAGN2017} also reported that IR-selected AGN lie on or above the MS. For AGN identified with optical spectroscopy, \citet{vietriTypeIIAGNhost2022} found a broader distribution of SFRs than the MS, similar to that of non-AGN galaxies: massive galaxies have a larger fraction of quenched galaxies, while lower stellar mass galaxies are on the MS. This picture is consistent with our SFR--M$_{\star}$ distribution for the SED AGN, which is the most similar to that of non-AGN. 
Our results and previous findings underscore an important caveat in AGN selection: the host galaxies of different AGN types systematically differ from one another. Thus, it is important to have control galaxies that match the AGN sample in stellar mass, redshift, and SFR.

\subsection{Control samples}\label{sect:controls}

We compared the merger and the AGN populations to control samples of non-mergers or non-AGN, respectively. Specifically, the controls satisfied the following conditions:
\begin{align}\label{control_eq}
    &|z_{control}-z_{sample}| \leq \Delta z \times z_{sample}\, ,\\
    &|\log \; \rm{M}_{\star, control} - \log \; \rm{M}_{\star, sample} | \leq \Delta \rm{M}_{\star} \, ,\\
    &|\log \; \rm{SFR}_{control} - \log \; \rm{SFR}_{sample}| \leq \Delta \rm{SFR} \, ,
\end{align}
where $\Delta z = 0.03$, $\Delta \log{\rm M}_{\star} = 0.1\,{\rm dex}$, and $\Delta \log {\rm SFR} = 0.3\,{\rm dex}$. We chose these values based on the photo$-z$ precision and the median errors for M$_{\star}$ and SFR. For each galaxy in the original sample, we required at least 10 counterparts that satisfied these criteria. We randomly picked 10 if more than 10 controls were found. If there were fewer than 10 controls, we iteratively increased the tolerance of each parameter by a factor of 1.5. This operation was performed up to a maximum of three times, otherwise, we rejected the galaxy. For non-AGN controls, we only considered galaxies that do not host any detected AGN. 

To make a fair comparison with the results presented in \citetalias{lamarcaDustPowerUnravelling2024a}, we built control samples without matching SFRs and ran our experiments with these controls. However, the results were qualitatively the same, with all variations within $1\sigma$ uncertainties. Therefore, we did not report the results of this test.

%--------------------------------------------------------------------

\subsection{Merger--AGN relation using a binary AGN classification}

% We ran two experiments assuming a binary AGN/non-AGN classification. 
In the first set of experiments, we investigate the merger--AGN connection with a binary AGN/non-AGN classification. 
If mergers trigger AGN, we would expect a higher incidence of AGN in mergers than in non-merger controls. If mergers are the primary trigger, we should observe that most AGN are in mergers. We divided the sample into two redshift bins, $z$-bin 1 $= [0.5; 1.25)$ and $z$-bin 2 $=[1.25;2.0]$, with roughly equal numbers of AGN. 

\subsubsection{AGN frequency in mergers and non-mergers}

\begin{figure}
    \centering
    \includegraphics[width=0.45\textwidth]{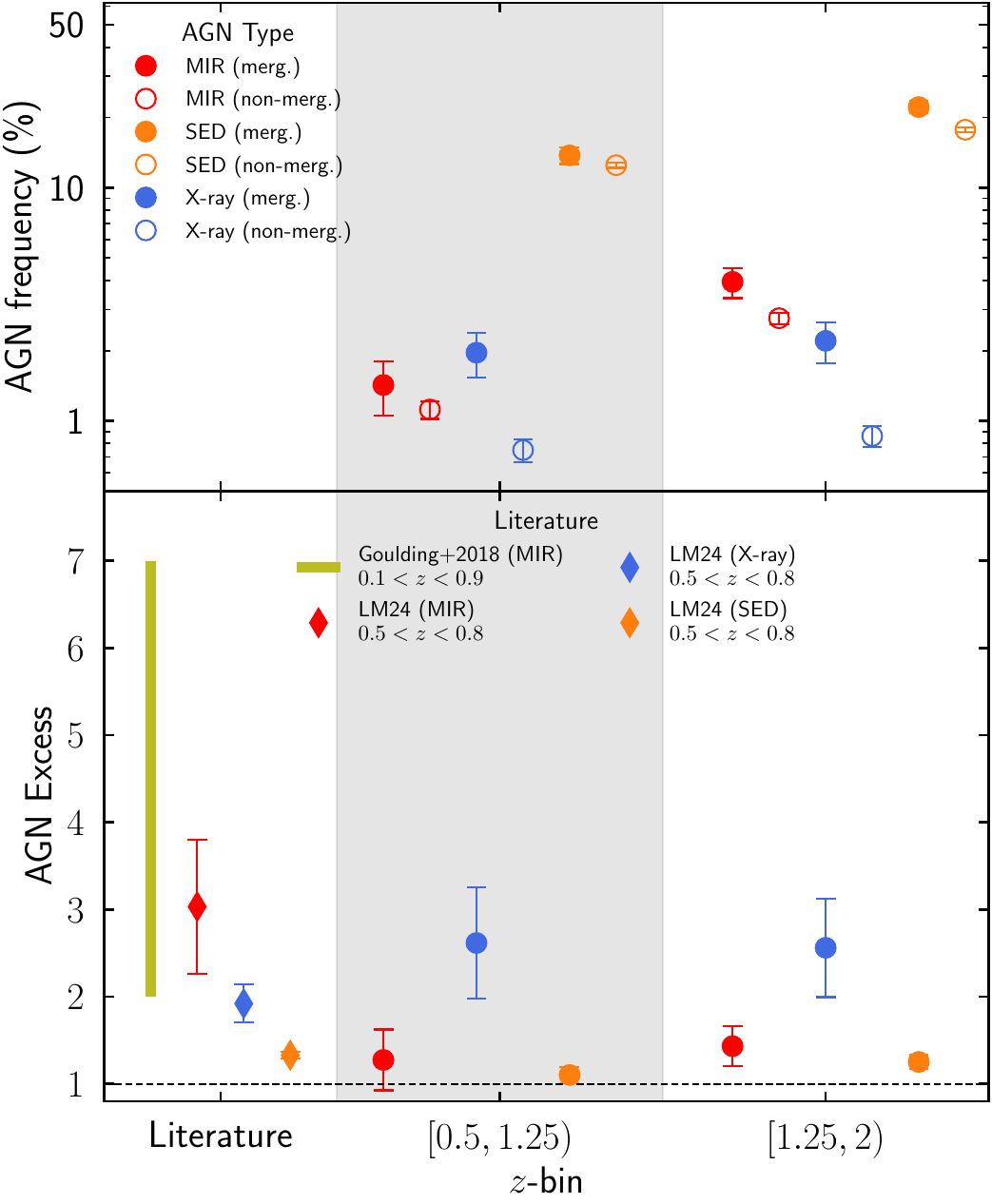}
    \caption{
    \textit{Top}: Frequency of AGN (MIR: red; X-ray: blue; SED: orange) in mergers, indicated with filled symbols, and non-merger controls, empty symbols. 
    \textit{Bottom}: AGN frequency in mergers divided by the AGN frequency in the relative non-mergers (i.e. AGN excess). 
    }
    \label{fig:agn_freq}
\end{figure}

We show the frequency of AGN in mergers and non-merger controls in Fig.~\ref{fig:agn_freq} and report the exact values in Appendix~\ref{app:tables}, Table~\ref{tab:agn_freq}. For all AGN types and $z$-bins, we observe a marginally higher frequency of AGN in mergers, demonstrating that mergers are a viable way to fuel accretion onto SMBHs. We also calculated the AGN excess by taking the ratio of the AGN frequency in mergers relative to non-mergers. 
About $1.5\%$ and $4\%$ of mergers host MIR AGN in $z$-bins 1 and 2, respectively, compared to $\sim 1\%$ and $3\%$ in non-mergers. Hence, the MIR AGN excess is $1.3-1.4$ in mergers, which is significant at $2\sigma$ level in $z$-bin 2 but is compatible with no excess in $z$-bin 1. 
Roughly $2\%$ of mergers host X-ray AGN in both $z$-bins, compared to $<1\%$ in non-mergers. This leads to an X-ray AGN excess ratio of 2.6 in mergers (significant at $\sim2.5\sigma$), which is the highest among the three AGN types. 
Around 14\% and 22\% of mergers host SED AGN in $z$-bins 1 and 2, respectively, compared to $\sim 12\%$ and $\sim 18\%$ in non-mergers. Consequently, the SED AGN excess is the lowest among the three types. The excess is almost negligible in $z$-bin 1 ($1.10\pm0.09$), and modest ($1.20\pm0.07$), though significant at $\sim 3 \sigma$, in the $z$-bin 2.

% Comparison with previous results. 
Recent studies at $z<1$ found higher MIR AGN excesses in mergers than what we observe \citep[excesses $\simeq 1.5$--$7$;][]{gouldingGalaxyInteractionsTrigger2018, bickleyAGNsPostmergersUltraviolet2023, lamarcaDustPowerUnravelling2024a}. Our X-ray AGN excess is in line with the results in \citet{lacknerLateStageGalaxyMergers2014}. On the other hand, the X-ray AGN excess we find is larger than what was observed at lower redshift by \citet[][excess $\simeq 1.8$]{bickleyAGNsPostmergersUltraviolet2023} and \citetalias[][excess $\simeq 1.9$]{lamarcaDustPowerUnravelling2024a}, but still within the $1\sigma$ uncertainty. Our results for SED AGN are slightly lower but still in agreement with previous studies of SED AGN \citepalias{lamarcaDustPowerUnravelling2024a} and of optical-emission-lines-selected AGN \citep{gaoMergersTriggerActive2020, tanakaGalaxyCruiseDeep2023}. Fewer studies investigated AGN frequency in mergers and non-mergers at $z>1$. Our results are in qualitative agreement with \citet{silvaGalaxyMergers252021}, who observed a mild excess of AGN in mergers that decreases as a function of redshift. 
This experiment shows that mergers can trigger AGN, but it is unclear if there is a strong dependence on redshift. 
There is some expectation \citep{kocevskiAreComptonthickAGNs2015} that mergers may play a less important role at intermediate and high redshifts and, consequently, the greater role that secular processes play in feeding SMBHs, due to the larger availability of cold gas \citep[][]{tacconiHighMolecularGas2010}. 
%As pointed out by \citet{kocevskiAreComptonthickAGNs2015}, the effect of major mergers might be diluted at higher $z$ compared to lower $z$, where they are necessary to bring fresh gas supplies to galaxies. 
%This experiment makes it clear that mergers can trigger AGN but also that they are not the only ones responsible. Moreover, the AGN excesses are lower at higher $z$ than at $z<1$. One possible explanation is the larger availability of cold gas at intermediate and high redshifts \citep[][]{tacconiHighMolecularGas2010} and, consequentially, the greater role that secular processes play in feeding SMBHs. As pointed out by \citet{kocevskiAreComptonthickAGNs2015}, the effect of major mergers might be diluted at higher $z$ compared to lower $z$, where they are necessary to bring fresh gas supplies to galaxies. 

% Much lower AGN excess for the MIR AGN compared to low redshift and previous studies. Possibilities: at high z galaxies simply are more dust-obscured and the effect of mergers is diluted; we create SFR-matched controls; WISE selected brighter galaxies compared to IRAC. 

\subsubsection{Merger fraction in AGN and non-AGN}

\begin{figure}%[h]  
    \centering\includegraphics[width=0.45\textwidth]{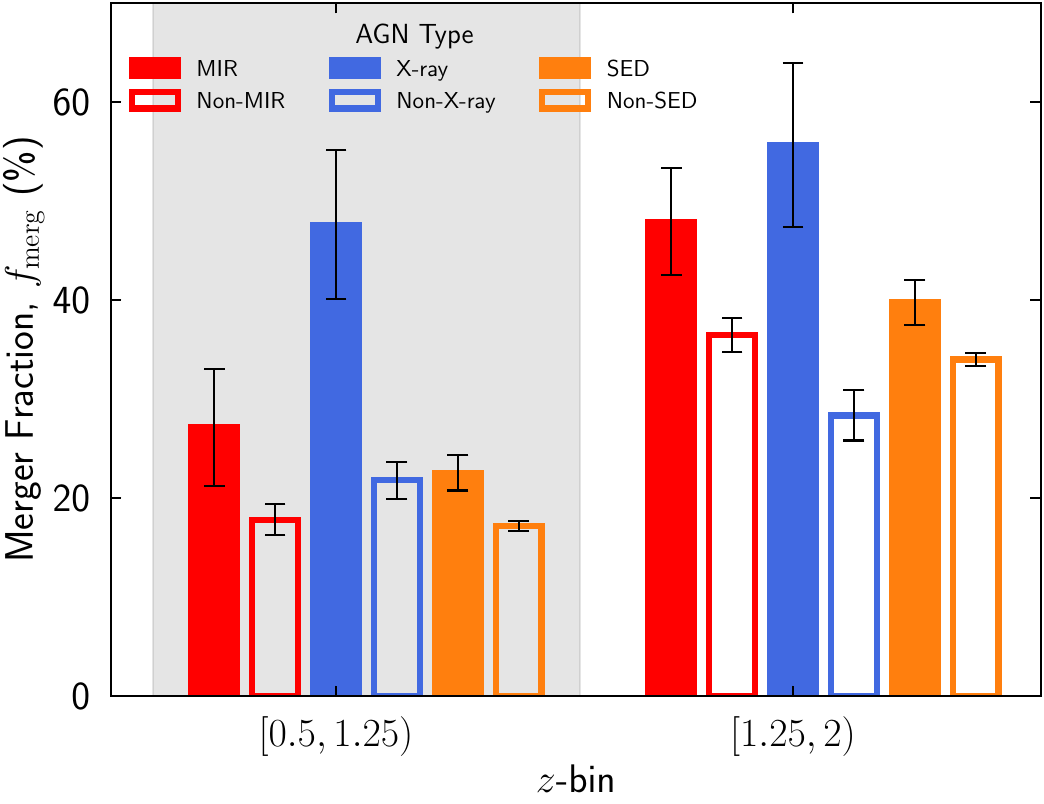}
    \caption{
    Merger fraction in the AGN (filled bars) and non-AGN controls (empty bars), with the same colour coding as Fig.~\ref{fig:agn_freq}. 
    }
    \label{fig:merg_frac}
\end{figure}

We report the merger fraction ($f_{\rm merg}$) in AGN and non-AGN controls in Fig.~\ref{fig:merg_frac} and Table \ref{tab:merg_frac}. For all $z$-bins and AGN types, $f_{\rm merg}$ is higher in AGN than in non-AGN controls, strengthening the merger--AGN connection. About half of the X-ray AGN in both $z$-bins are hosted by mergers ($48\%$ and $56\%$), while $f_{\rm merg}$ is $22\%$ and $28\%$ in the corresponding non-AGN controls. MIR AGN are hosted by mergers in $27\%$ ($48\%$) of the cases in $z$-bin 1 ($z$-bin 2), while $18\%$ ($36\%$) of the non-MIR-AGN controls are mergers. For the SED AGN, $f_{\rm merg}$ is around 23\% (40\%) in $z$-bin 1 ($z$-bin 2), while $f_{\rm merg}$ is around 17\% (34\%) in the non-SED-AGN controls. Merger fractions are larger in the second $z$-bin for all AGN types, in agreement with the previously observed rising $f_{\rm merg}$ with increasing $z$ \citep[][]{ferreiraGalaxyMergerRates2020}.
%We conclude that mergers are not the main trigger of AGN activity in most cases. 

Overall, we report only a slight enhancement of $f_{\rm merg}$ in AGN compared to non-AGN controls.
$f_{\rm merg}$ is $\sim50\%$ only for the X-ray AGN in both $z$-bins and MIR AGN in $z$-bin 2, indicating that mergers might be the main triggering mechanism or at least play an important role in these cases. However, we also observe significant merger fractions in non-AGN controls. Hence, AGN could be ubiquitous in mergers and non-mergers but episodic. We should consider a scenario in which not all mergers trigger the AGN phase. In this case, the difference between $f_{\rm merg}$ in AGN and $f_{\rm merg}$ in non-AGN represents the real fraction of mergers triggering AGN. For example, considering MIR AGN in $z$-bin 2, only $48\%-36\%=12\%$ of all MIR AGN are triggered solely by mergers. In a different case, it might be possible that all mergers trigger AGN, but the AGN happen to be ``off'' in some galaxies, given the mismatch in the merger and AGN timescales. This scenario would imply that $48\%+36\%=84\%$ of MIR AGN are triggered by mergers.

Our results are in qualitative agreement with previous studies.  \citet{mechtleyMostMassiveBlack2016} studied a sample of quasars up to $z=2$ and observed a marginally larger $f_{\rm merg}$ in quasar hosts than in inactive galaxies. Similarly, \citet{marianMajorMergersAre2019} and \citet{kocevskiCANDELSConstrainingAGNMerger2012} found only a mildly larger $f_{merg}$ in X-ray AGN compared to inactive control galaxies. \citet{villforthCompleteCatalogueMerger2023} performed a review and concluded that in most cases, $f_{\rm merg}$ in AGN is consistent with no excess over controls. Nevertheless, other studies reported higher $f_{\rm merg}$. \citet{fanMostLuminousHeavily2016} found that the most luminous dust-obscured AGN are more likely to be classified as disturbed, with a high $f_{\rm merg}$ (62\%). With a larger sample including faint sources, \citet{bonaventuraRelationAGNHostgalaxy2025} observed that a high fraction of AGN hosts are strongly disturbed. Likewise, \citet{donleyEvidenceMergerdrivenGrowth2018} found that more than 70\% of MIR AGN are classified either as interacting or highly disturbed. However, these studies are biased toward bright and heavily obscured objects, which are more supported observationally to be connected with major mergers \citep{ricciGrowingSupermassiveBlack2017, ricciHardXrayView2021}.

%--------------------------------------------------------------------

\subsection{The merger link with AGN relative and absolute power}

We now examine the merger--AGN connection using a continuous approach.%, trying to answer the question: does the role of mergers change as a function of AGN properties? 
We make use of the relative and absolute AGN power, traced by the AGN fraction ($f_{\rm AGN}$) and the AGN accretion disc luminosity (L$_{disc}$), respectively.

\subsubsection{The merger fraction and AGN fraction relation}

\begin{figure}
    \centering
    % \sidecaption
    \includegraphics[width=0.49\textwidth]{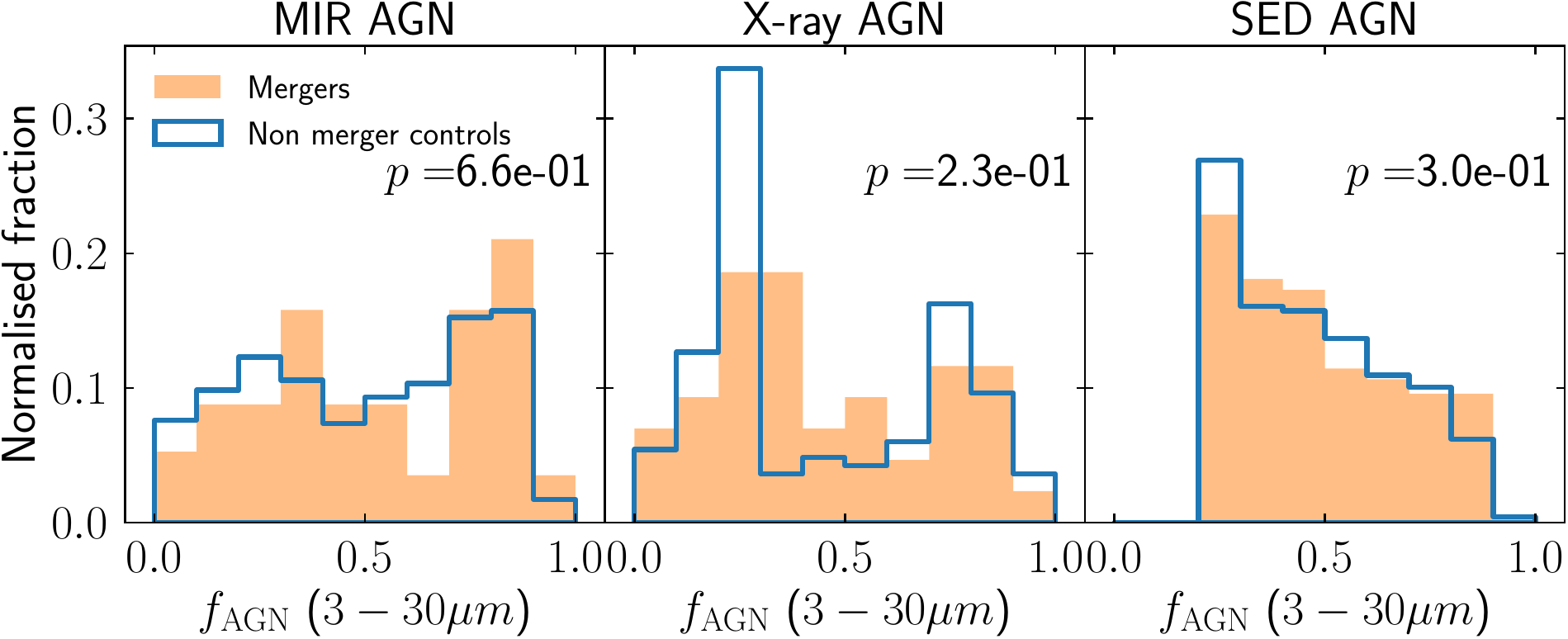}
    \caption{
    AGN fraction normalised distributions for mergers (orange) and non-merger controls (empty blue bars), divided by AGN type. In each panel, we display the results of the KS-test run for the two populations.}
    \label{fig:fagn_dist}
\end{figure}

We display the $f_{\rm AGN}$ distribution for mergers and non-merger controls in Fig.~\ref{fig:fagn_dist}. MIR AGN show large fractions of host galaxies with $f_{\rm AGN}>0.7$ for both mergers and non-mergers. This is expected because $f_{\rm AGN}$ is measured in the $3-30\,\mu$m range, and the MIR diagnostics preferentially trace obscured AGN. However, mergers harbouring MIR AGN show a larger fraction at $f_{\rm AGN}>0.7$ compared to non-mergers. %, while the opposite is true for lower $f_{\rm AGN}$ values. 
For the X-ray AGN, the $f_{\rm AGN}$ distributions appear to be bimodal, with peaks at $0.3$ and $0.8$ for both mergers and non-mergers. Non-mergers have larger fractions of galaxies with $f_{\rm AGN}\leq0.3$ than mergers. Finally, the $f_{\rm AGN}$ distributions for the SED AGN do not show a significant difference between mergers and non-mergers, only a mildly larger fraction of mergers at $f_{\rm AGN}\geq 0.8$. We ran two-sample Kolmogorov-Smirnov \citep[KS;][]{hodgesSignificanceProbabilitySmirnov1958} tests to measure the statistical difference between the distributions in mergers and non-mergers (reported in Fig.~\ref{fig:fagn_dist}), which confirm that the mild differences are not statistically significant.

\begin{figure}
    \centering
    \includegraphics[width=.45\textwidth]{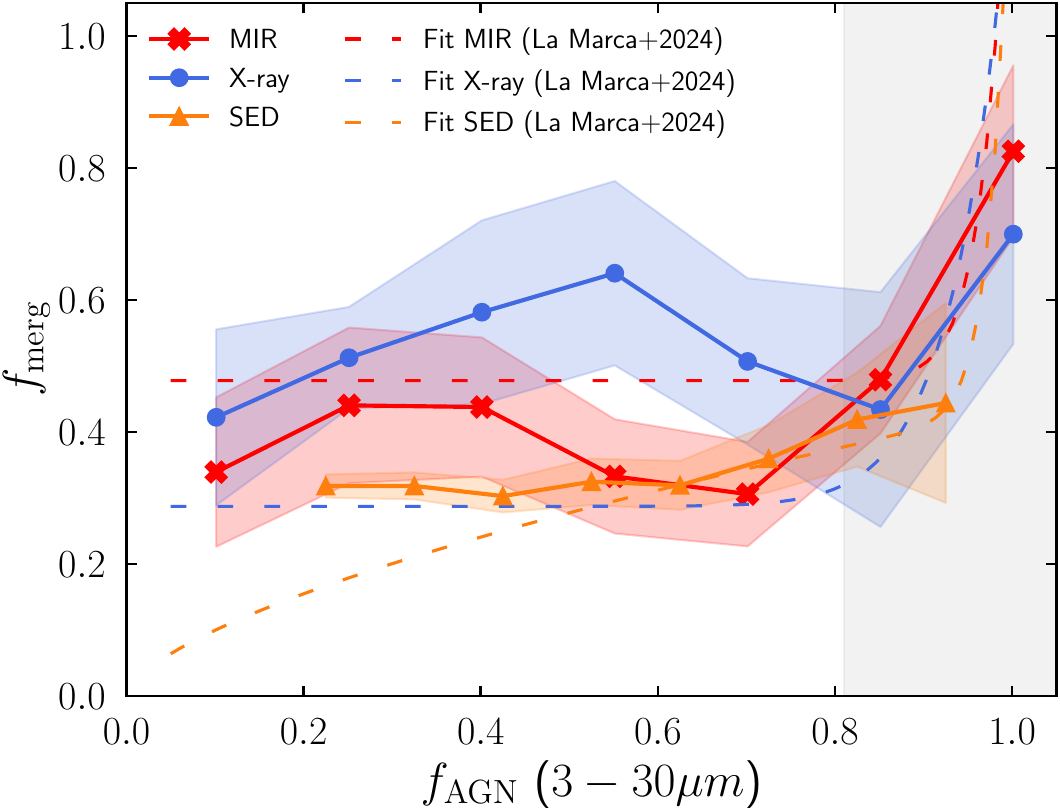}
    \caption{
    Merger fraction as a function of AGN fraction for the MIR AGN (red crosses), X-ray AGN (blue circles) and SED AGN (orange triangles). The solid lines and coloured region show the running median and standard deviation of each relation. The dashed lines indicate the fits to the relations in \citetalias{lamarcaDustPowerUnravelling2024a}. %The same colours represent the same AGN types. 
    The light grey area indicates the region where the relationships' behaviours change strongly (see text).
    }
    \label{fig:fmerg_fagn}
\end{figure}

We present the $f_{\rm merg}$--$f_{\rm AGN}$ relationship for the three AGN types in Fig.~\ref{fig:fmerg_fagn}. We computed $f_{\rm merg}$ in $N$ bins (with $N$ randomly sampled between 6 and 20) of $f_{\rm{AGN}}$, equally spaced in the range $0$--$1$. 
%Bootstrapping with resampling is used (1000 samples for each population). 
The trends reported represent the running medians of 1000 bootstrap samples for each population. The MIR and SED AGN show a rather flat $f_{\rm merg}$ in the range $f_{\rm AGN}=0.1$--$0.6$. At $f_{\rm AGN}>0.6$, MIR AGN have a rise in $f_{\rm merg}$ while SED AGN have a more gentle increase in $f_{\rm merg}$.
X-ray AGN show a more complicated trend: a mildly increasing $f_{\rm merg}$ until $f_{\rm AGN}=0.6$, followed by a mild decrease in $f_{\rm merg}$ in the range $f_{\rm AGN}=0.6$--$0.8$, in turn followed by an increasing $f_{\rm merg}$ at $f_{\rm AGN}>0.8$. Moreover, half of the X-ray AGN inhabit mergers. %, since $f_{\rm merg}\simeq50\%$ across almost the entire $f_{\rm AGN}$ range. 
Based on these results, we conclude that mergers could be the main triggering mechanisms for the most dominant MIR and X-ray AGN ($f_{\rm AGN}>0.8$). 

We compare the $f_{\rm merg}$--$f_{\rm AGN}$ relation to the same relation in \citetalias{lamarcaDustPowerUnravelling2024a} at $z<0.8$. The low-redshift relation revealed two different regimes for all AGN types: a flat $f_{\rm merg}$ up to $f_{\rm AGN}<0.8$, and a subsequent steep rise of $f_{\rm merg}$ at $f_{\rm AGN}> 0.8$. To compare with the relation in COSMOS-Web, we parametrised the \citetalias{lamarcaDustPowerUnravelling2024a} $f_{\rm merg}$--$f_{\rm AGN}$ relation, with details on the fitting method and results in Appendix~\ref{app:fit}. In Fig.~\ref{fig:fmerg_fagn}, we report the fitted relations per AGN type. Considering the $f_{\rm AGN}<0.8$ regime, there is a qualitatively good agreement between our and \citetalias{lamarcaDustPowerUnravelling2024a} results, with a mild difference. The plateau level for each AGN selection is slightly different at low and high redshifts. Considering the same SED AGN definition adopted here ($f_{\rm AGN}>0.2$), SED AGN have on average the same $f_{\rm merg}$ plateau value shown in \citetalias{lamarcaDustPowerUnravelling2024a} but show a flatter trend. MIR AGN show a larger merger fraction at $z<0.8$ than at $0.5<z<2$, possibly due to the different luminosity ranges covered. The contrary is true for the X-ray AGN, which are more frequently hosted by mergers at higher redshifts. Nevertheless, given the large uncertainties, our findings are consistent with no significant difference from the trend reported in \citetalias{lamarcaDustPowerUnravelling2024a}. Although similar in the low-dominance AGN regime, more differences emerge at $f_{\rm AGN}>0.8$. MIR, X-ray, and SED AGN show a steeper rise at low redshift compared to high redshift. This is most evident for the SED AGN, where there is only a mild increase for $0.5<z<2$ galaxies.

Two aspects require further discussion. First, the reason why SED AGN do not show a steep rising $f_{\rm merg}$ for dominant AGN may be because SED AGN are usually fainter than MIR and X-ray AGN. Even relatively dominant SED AGN with $f_{\rm AGN}>0.8$ might still be faint in absolute terms. Lower luminosities correspond to lower SMBH accretion rates, which secular processes could sustain if enough gas is available. 
%Most SED AGN inhabit galaxies at $1<z<2$, corresponding to the peak of star-formation activity and gas content in galaxies \citep{madauCosmicStarFormationHistory2014}, so mergers might be only a secondary fuelling mechanism even for dominant SED AGN. 
Second, $f_{\rm merg}$ shows a steeper rise for dominant MIR and X-ray AGN at low $z$ than at high $z$, hinting at a lower importance of mergers in triggering dominant AGN at cosmic noon than at lower redshifts. This may be possible, given the different gas supplies available. At higher $z$, galaxies have larger gas fractions and stochastic fuelling by internal processes could play a greater role than at lower $z$ \citep{kocevskiCANDELSConstrainingAGNMerger2012, kocevskiAreComptonthickAGNs2015}. At $z<1$, galaxies generally have lower gas fractions, and so external sources are necessary to enlarge the gas content and fuel dominant AGN \citep{Tang2025ALMAObservations}.

Nevertheless, it is worth noting that the availability of gas alone is not sufficient to trigger SMBH accretion. Efficient accretion also requires the removal of angular momentum, which can be achieved not only via major mergers but also bars or secular instabilities, particularly in gas-rich disks \citep[][]{combesFuelingAGN2001}. Hence, a key difference might be that mergers at low-$z$ are necessary not only for removing angular momentum but also to bring new cold gas to fuel nuclear activity. Thus, it is likely that the role of mergers, as a function of $f_{\rm AGN}$, evolves with cosmic time (and/or gas content): mergers gain importance in fuelling dominant AGN with decreasing $z$ (or gas content), becoming the sole viable mechanism to feed the most dominant AGN at $z=0$.

\subsubsection{The merger fraction and L$_{disc}$ relation}

\begin{figure}
    \centering
    % \sidecaption
    \includegraphics[width=0.49\textwidth]{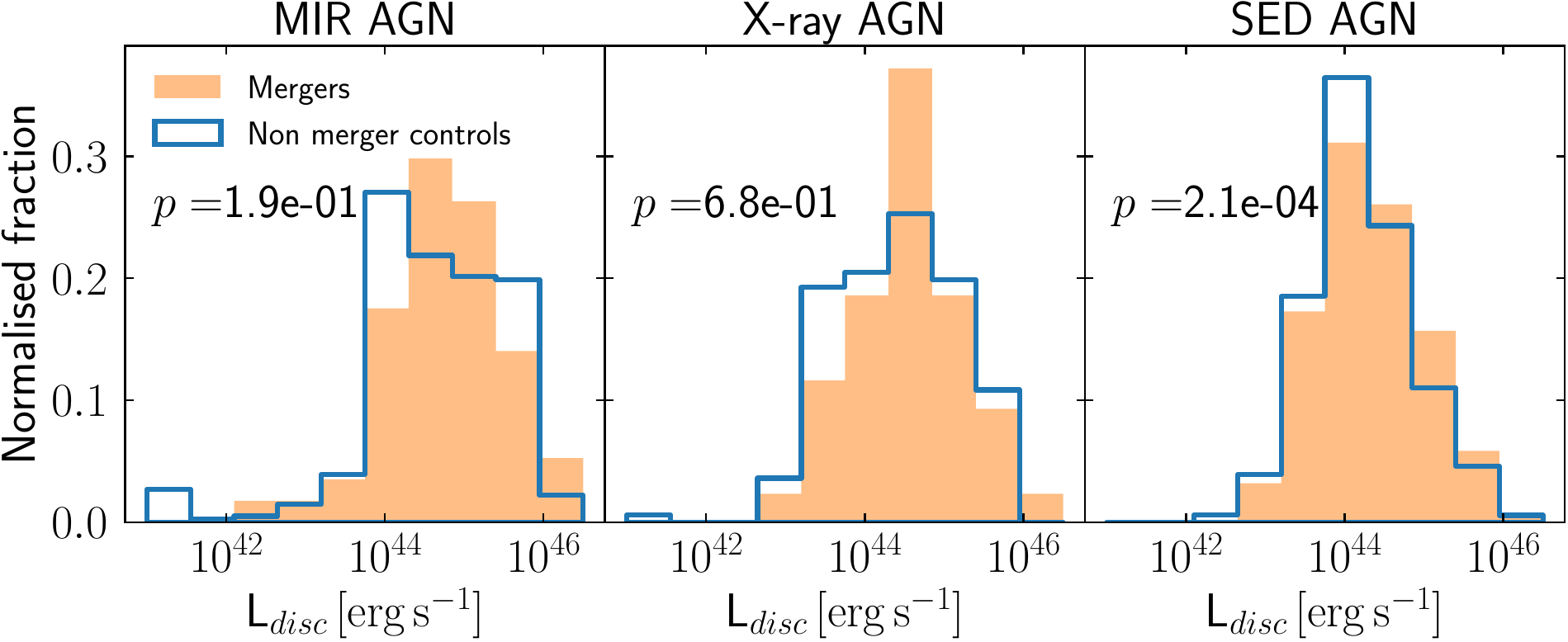}
    \caption{
    AGN luminosity, $\rm{L}_{disc}$, normalised distributions divided by AGN type, as in Fig.~\ref{fig:fagn_dist}.}
    \label{fig:Lagn_dist}
\end{figure}

\begin{figure}
    \centering
    \includegraphics[width=0.49\textwidth]{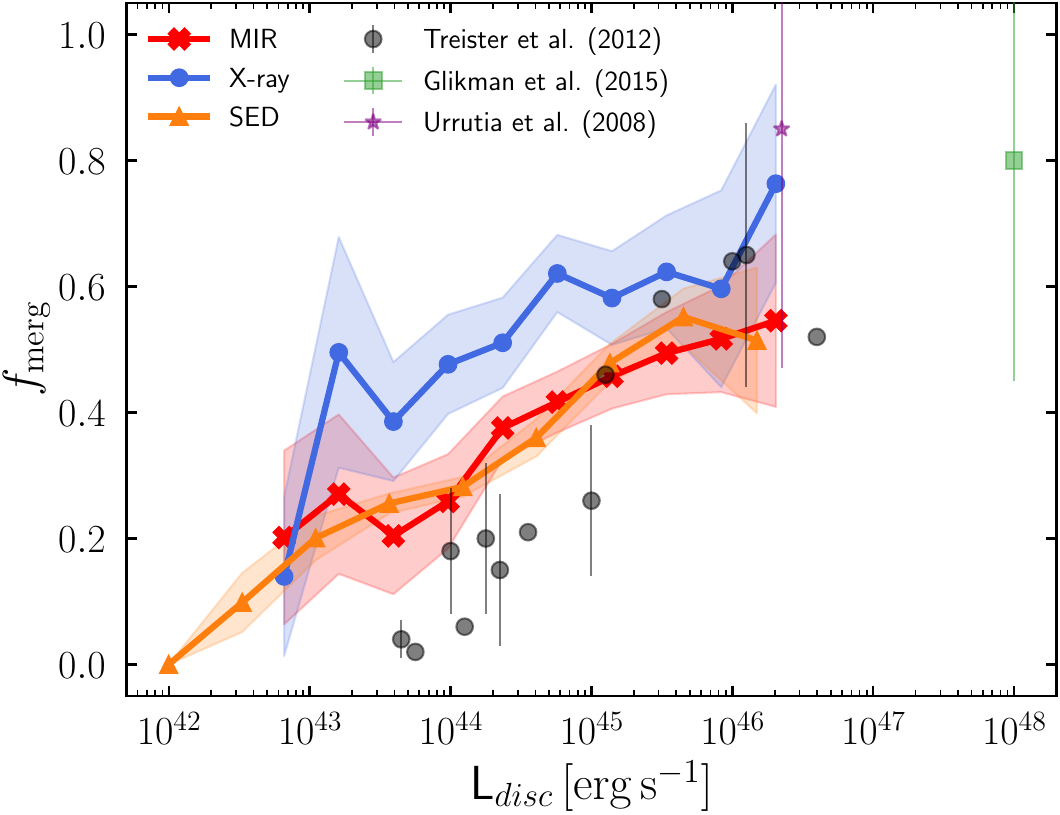}
    \caption{
    The merger fraction--AGN luminosity relation for the three AGN types. The solid line and coloured region show the running median and standard deviation of each relation. Previous literature findings are reported.}
    \label{fig:fmerg_Lagn}
\end{figure}

Now we analyse the merger--AGN connection using the absolute AGN power. We show the normalised L$_{disc}$ distribution for mergers and non-merger controls in Fig.~\ref{fig:Lagn_dist}. Overall, the merger and non-merger distributions are similar for all AGN types, although, in the case of SED AGN, mergers show larger fractions of bright AGN (L$_{disc}$) compared to non-merger controls, as confirmed by the KS-test results. 
It is worth mentioning that, looking at the very bright end (L$_{disc}\gtrsim10^{46}\,{\rm erg\,s^{-1}}$), there is an excess in the fraction of mergers compared to the fraction of non-mergers also for MIR and X-ray AGN.

We plot the $f_{\rm merg}$--L$_{disc}$ relation in Fig.~\ref{fig:fmerg_Lagn}, in the same way as for the $f_{\rm merg}$--$f_{\rm AGN}$ relation, with the only difference being the $N$ L$_{disc}$ bins evenly spaced on a logarithmic scale. For all three AGN types, $f_{\rm merg}$ rises as a function of increasing L$_{disc}$, with mergers accounting only for $\lesssim 20\%$ of the faint AGN (L$_{disc} \lesssim 10^{43}\,{\rm erg\,s^{-1}}$) but being the dominant trigger of the brightest AGN (L$_{disc}\gtrsim10^{46}\,{\rm erg\,s^{-1}}$). These results imply that the role of mergers is directly linked to the absolute AGN power, with mergers being the most efficient process to fuel the most powerful AGN. We could also interpret these results from another angle, i.e., secular processes become less efficient in fuelling AGN activity above a certain threshold.
% or, in other words, their efficiency in feeding accretion onto SMBHs decreases with increasing L$_{disc}$. 

These findings align with most previous results in the literature. The early work by \citet{urrutiaEvidenceQuasarActivity2008} analysed a limited sample of bright quasars and found that $>80\%$ of them exhibit clear signs of recent or ongoing major interactions. Similarly, \citet{glikmanMajorMergersHost2015} observed an $f_{\rm merg}$ of $80\%$ for a sample of extremely bright quasars. \citet{treisterMajorGalaxyMergers2012, donleyEvidenceMergerdrivenGrowth2018} and \citet{euclidcollaborationEuclidQuickData2025} also reported an increasing $f_{\rm merg}$ as a function of the AGN luminosity, with mergers becoming the dominant mechanism for L$_{\rm AGN}>10^{45.5}\,{\rm erg\,s^{-1}}$. 
Nonetheless, other studies found no clear correlation between $f_{\rm merg}$ and AGN luminosity. For instance, our results are at odds with those reported by \citet{allevatoXMMNewtonWideField2011}, \citet{hewlettRedshiftEvolutionMajor2017}, and \citet{villforthCompleteCatalogueMerger2023}, which found no evidence to support the scenario where the most luminous AGN reside mostly in mergers. However, the reasons for this conflict remain unclear. It may be due to the small sample sizes and/or the different selections of those studies, which makes a fair comparison difficult. 
%Both limitations are, in principle, controllable with large area space observations. 

A further impediment in studying the merger--AGN connection is the timescales. While the AGN duty cycle usually lasts $10-100\,{\rm Myr}$ \citep{marconiLocalSupermassiveBlack2004}, major interactions happen over a wider range, from a few hundred Myr up to a few Gyr. Thus, even if all major mergers trigger AGN, the AGN might be in an ``off'' status when observed. Bearing this complication, better datasets are necessary to obtain more accurate results, but there will always be intrinsic uncertainty in quantifying the merger--AGN link.

%--------------------------------------------------------------------
\section{Summary and Conclusions}\label{sect:Conclusions}

In this paper, we examined the merger--AGN connection at $0.5\leq z \leq 2$ in the COSMOS-Web field. We selected a stellar-mass-limited sample with rich multi-wavelength data and then performed detailed SED decomposition. We identified AGN with three different diagnostics, MIR colours, X-ray detections, and SED modelling, and detected major mergers with the use of CNNs trained on mock observations from two independent cosmological galaxy formation simulations. We examined the merger--AGN relation first in terms of a binary active/non-active classification and then in terms of absolute and relative AGN power, using the AGN fraction ($f_{\rm AGN}$) and AGN accretion disc luminosity (L$_{disc}$), respectively. Our main results are:
\begin{enumerate}

    \item A moderate AGN excess in mergers compared to non-mergers for all AGN types, with the largest excess measured for X-ray AGN (a factor of 2.6), and the lowest for SED AGN. Thus, mergers can trigger AGN activity also out to $z=2$. Although these excesses are slightly lower than at lower redshifts, it is unclear if there is a strong redshift dependence. 
    
    \item A relatively larger merger fraction in AGN host galaxies compared to non-AGN controls for all AGN selections. However, these results do not allow for robust conclusions about the exact role of mergers in triggering AGN. 
    
    \item The $f_{\rm merg}$--$f_{\rm AGN}$ relation shows two regimes: a flat $f_{\rm merg}$ up to $f_{\rm AGN}\approx0.6$ for SED and MIR AGN, followed by a rise in $f_{\rm merg}$ at $f_{\rm AGN}>0.6$ in the case of MIR AGN, and a more gently increase in the case of SED AGN. 
    X-ray AGN show a more complicated trend, still exhibiting an increase of $f_{\rm merg}$ at $f_{\rm AGN}>0.8$.
    While most certainly mergers are only a secondary fuelling mechanism of relatively less dominant AGN, major mergers might be the principal channel to fuel nuclear activity in the most dominant AGN. 
    
    \item The $f_{\rm merg}$--L$_{disc}$ relation shows a monotonic trend for all AGN selections: $f_{\rm merg}$ continues to increase as a function of increasing AGN disc luminosity. Mergers appear as the main fuelling mechanism of extremely bright AGN, independent of the selection method. 
    
\end{enumerate}

To conclude, our results confirm that major mergers are a viable path to trigger AGN and show that mergers are strongly connected to the most luminous and dominant AGN, although possibly to a lesser degree compared to previous results at $z<1$. 
%These findings could be explained by the larger availability of cold gas at $z\simeq 1$--$2$, which could sustain intense AGN activity without the need of external gas sources. 
Major mergers still play a key role in fuelling the most powerful AGN, as secular processes might not be efficient in transporting large amounts of gas toward the central nuclei. 
To further confirm such a scenario, it is of pivotal importance to analyse larger samples of mergers and AGN to get more accurate measurements. In this direction, the planned observations of the European Space Agency survey mission \textit{Euclid} \citep{euclidcollaborationEuclidOverviewEuclid2024} will allow a significant extension of this work over a larger area and pinpoint the exact role of mergers in triggering AGN \citep{euclidcollaborationEuclidQuickData2025}. Moreover, future studies should focus on distinguishing pre- and post-mergers to understand which stages are more likely responsible for fuelling nuclear activity and to which AGN phase they are connected.

% {\color{red} Lingyu - Future outlook, pushing to even higher z with JWST, Euclid, merger stages and connection with different AGN types. Could compare with studies which looked at the correlation or not between AGN luminosity and SFR of host galaxies. These studies generally find that at very high AGN luminosities, there is a correlation between AGN luminosity and SFR. However, at low and moderate AGN luminosities, there is no such correlation. A possible interpretation is that mergers fuel the SMBH growth and bulge growth for high-luminosity AGN (i.e. establishing a direct link) and secular processes fuel the less extreme AGN.}

%--------------------------------------------------------------------

\begin{acknowledgements}
This publication is part of the project `Clash of the titans:
deciphering the enigmatic role of cosmic collisions' (with project number VI.Vidi.193.113) of the research programme Vidi which is (partly) financed by the Dutch Research Council (NWO).

The training and testing of the CNN were carried out on the Dutch National Supercomputer (Snellius).
We thank SURF (www.surf.nl) for the support in using the National Supercomputer Snellius.

We thank the Center for Information Technology of the University of Groningen for support and access to the H\'abr\'ok high performance computing cluster.

\end{acknowledgements}

\bibliography{Merger-AGN_COSMOS_refs}

%--------------------------------------------------------------------

\begin{appendix}

\section{\texttt{CIGALE} parameters and reliability}\label{app:CIGALE}

We report in Table~\ref{tab:cigale_parameters} the complete configuration of \texttt{CIGALE} employed in this work. For each module, we provide the grid of parameters used to perform the SED decomposition. We used the default \texttt{CIGALE} values for the parameters not specified in Table~\ref{tab:cigale_parameters}. 

\begin{table*}[ht]
\small
% \scriptsize
\caption{Fitting parameters in the \texttt{CIGALE} runs.}
  \centering
  \begin{tabular}{lll}
  \hline\hline\\[-7pt]
  \multicolumn{3}{c}{\textbf{\large Star-Formation History}} \\
  \hline\\[-7pt]
  delayed $\tau$+starburst & e-folding time of the main population & 200, 500, 700, 1000--5000 (step 1000) Myrs \\
  & age the main population & 1000, 1500, $2000-8000$ (step 1000) Myrs \\
  & e-folding time of the late starburst population & 9000, 13000 Myrs \\
  & age of the late starburst population & 1, 50, 150 Myrs \\
  & mass fraction of the late starburst population & 0.0, 0.1, 0.2, 0.3 \\
  \hline\\[-7pt]
  \multicolumn{3}{c}{\textbf{\large Single Stellar Population}}\\
  \hline\\[-7pt]
  \citet{bruzualStellarPopulationSynthesis2003} & IMF & Chabrier (1) \\
  & metallicity & solar (0.02) \\
  \hline\\[-7pt]
  \multicolumn{3}{c}{\textbf{\large Dust attenuation}}\\
  \hline\\[-7pt]
  modified & V-band attenuation in the interstellar medium & 0.1, 0.5, 1, 1.5, 2, 2.5, 3, 4\\
  \citet{charlotSimpleModelAbsorption2000} & Av$_{ISM} $/ Av$_{BC}$+Av$_{ISM}$ & 0.25, 0.5, 0.75 \\
  & Power law slope of the attenuation in the ISM & $-0.7$ \\
  & Power law slope of the attenuation in the birth clouds & $-0.7$ \\
  \hline\\[-7pt]
  \multicolumn{3}{c}{\textbf{\large Dust emission}}\\
  \hline\\[-7pt]
  \citet{draineAndromedasDust2014} & Mass fraction of PAH & 0.47, 1.12, 2.5 \\
  & minimum radiation field & 5, 10, 25\\
  & power-law slope $\alpha$ in dM/dU $\propto U^{-\alpha}$ & 2.0 \\
  \hline\\[-7pt]
  \multicolumn{3}{c}{\textbf{\large X-ray}}\\
  \hline\\[-7pt]
  X-\texttt{CIGALE} X-ray & AGN photon index $\Gamma$ & 1.8 \\
  \citep{yangXCIGALEFittingAGN2020, yangFittingAGNGalaxy2022} & power slope $\alpha_{ox}$ & -1.8, -1.6, -1.4 \\
  & Max deviation from the $\alpha_{ox}-L_{\nu,2500 \AA }$ relation & 0.4 \\
  & AGN X-ray angle coefficients ($a_1$, $a_2$) & (0.5, 0)\\
  \hline\\[-7pt]
  \multicolumn{3}{c}{\textbf{\large AGN template}}\\
  \hline\\[-7pt]
  SKIRTOR & Average edge-on optical depth at 9.7 $\mu$m & 3, 7, 11 \\
  \citep{stalevski3DRadiativeTransfer2012} & torus density radial parameter $p$ & 1 \\
  & torus density angular parameter $q$ & 1\\
  & Angle between the equatorial plane and edge of the torus & 40$\degree$ \\
  & viewing angle & 30$\degree$ (type 1), 70$\degree$ (type 2) \\
  & AGN fraction, $f_{\rm AGN}$ & 0, 0.1, 0.2, 0.3, 0.45, 0.60, 0.75, 0.9, 0.99 \\
  & rest-frame wavelength range where $f_{\rm AGN}$ is computed & $3-30\, \mu$m 
  \\
  & extinction law of polar dust & SMC (0) \\
  & E(B-V) of polar dust & 0, 0.2, 0.4\\
  & temperature of polar dust & 100\\
  & emissivity of polar dust & 1.6\\
  \hline
  \end{tabular}
  
  \label{tab:cigale_parameters}
\end{table*}

\begin{figure*}
    \centering
    \includegraphics[width=.9\textwidth]{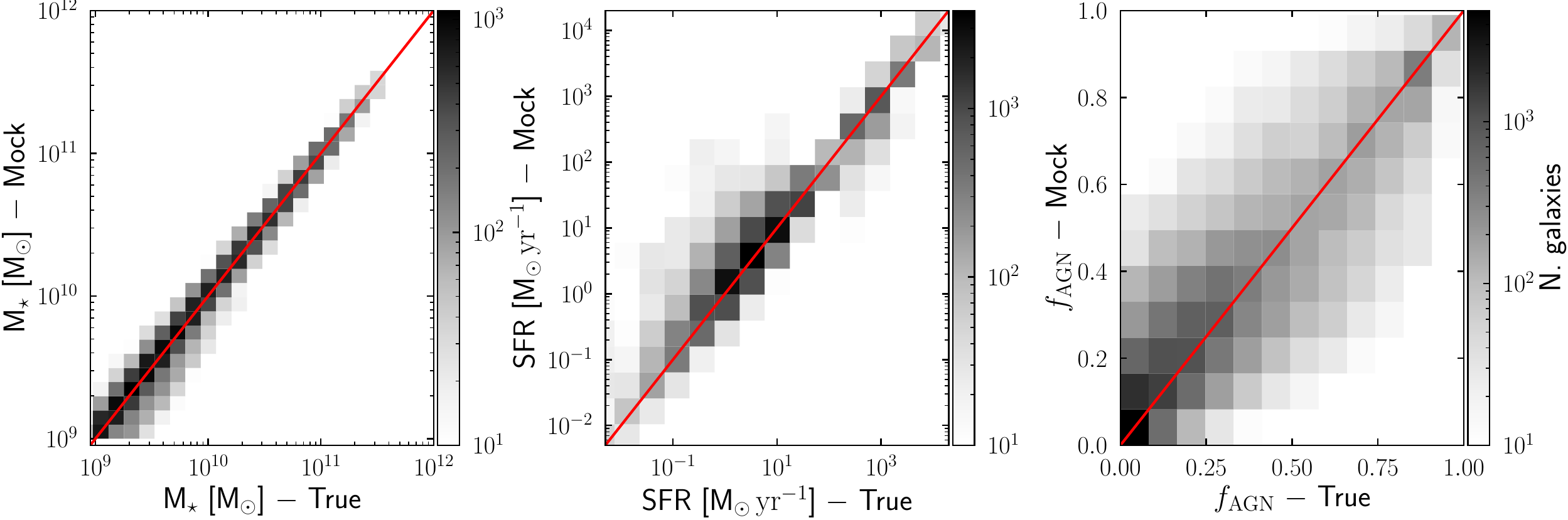}
    \caption{Comparison of the M$_{\star}$, SFR, and $f_{\rm AGN}$ measurements (\textit{left}, \textit{middle}, and \textit{right panel}, respectively) for the sources in the mock catalogue with the true values. The red lines indicate the 1-to-1 relations. 
    %\texttt{CIGALE} accurately recovers the true values of the mock sources for each quantity.
    }
    \label{fig:mock_analysis}
\end{figure*}

%\begin{figure}
 %   \centering
    %\includegraphics[width=.49\textwidth]{Plots_paper/Mock_analysis_gridplot.pdf}
    %\caption{Distributions of the deviations from the true values of M$_{\star}$, SFR, and $f_{\rm AGN}$ as a function of each other. The plots reveal no correlation between the measurements of the three properties.}
    %\label{fig:mock_analysis_grid}
%\end{figure}

\begin{figure*}
    \centering
    \includegraphics[width=.33\textwidth]{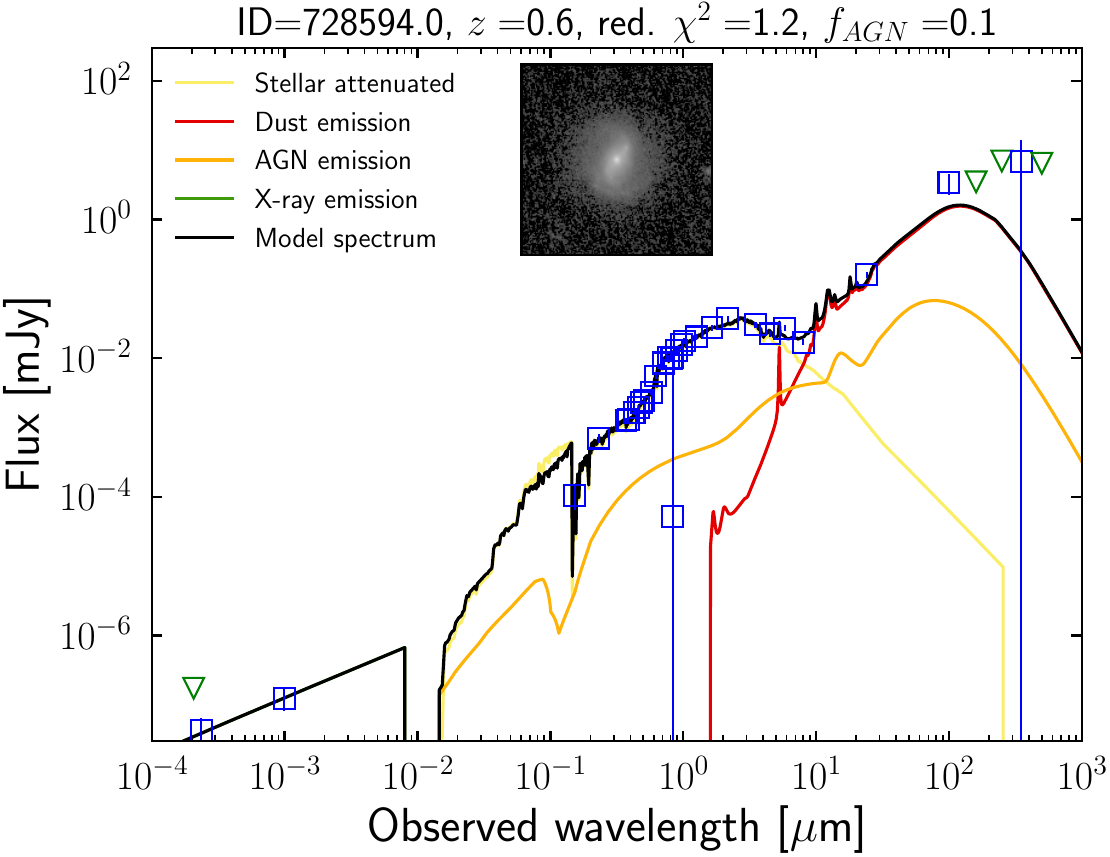}
    \includegraphics[width=.33\textwidth]{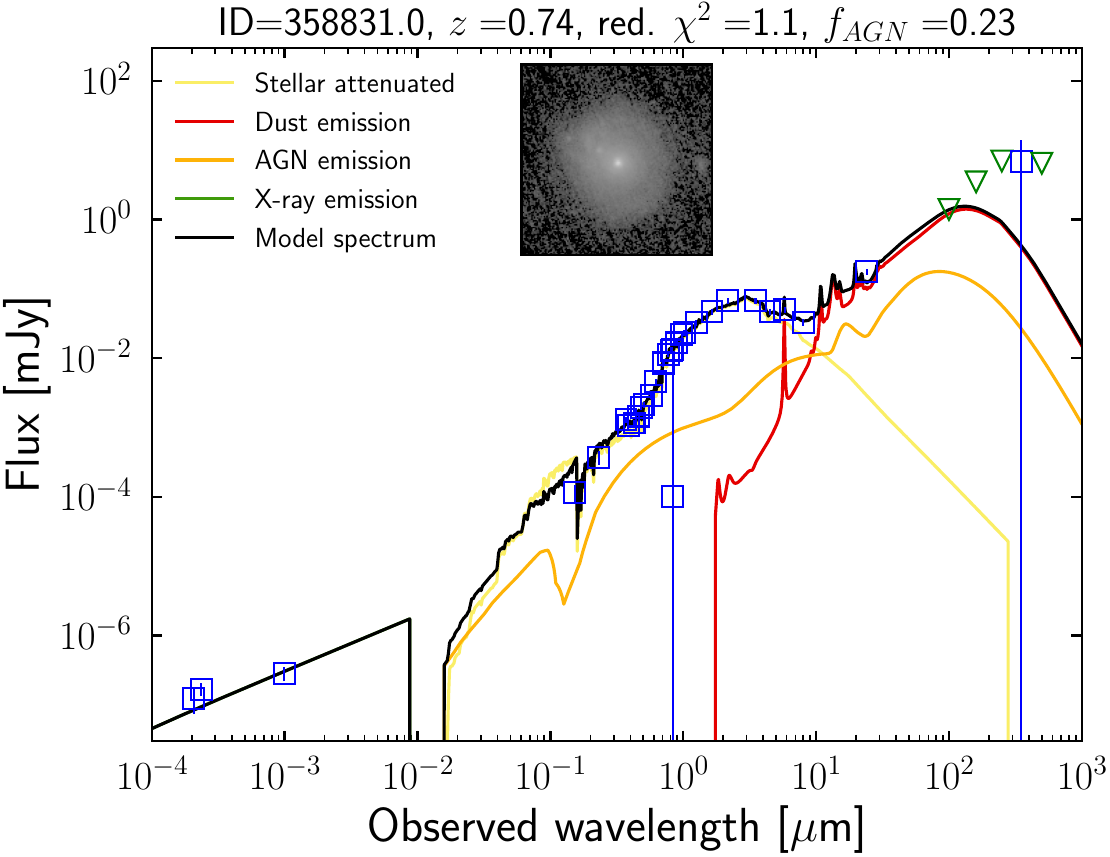}
    \includegraphics[width=.33\textwidth]{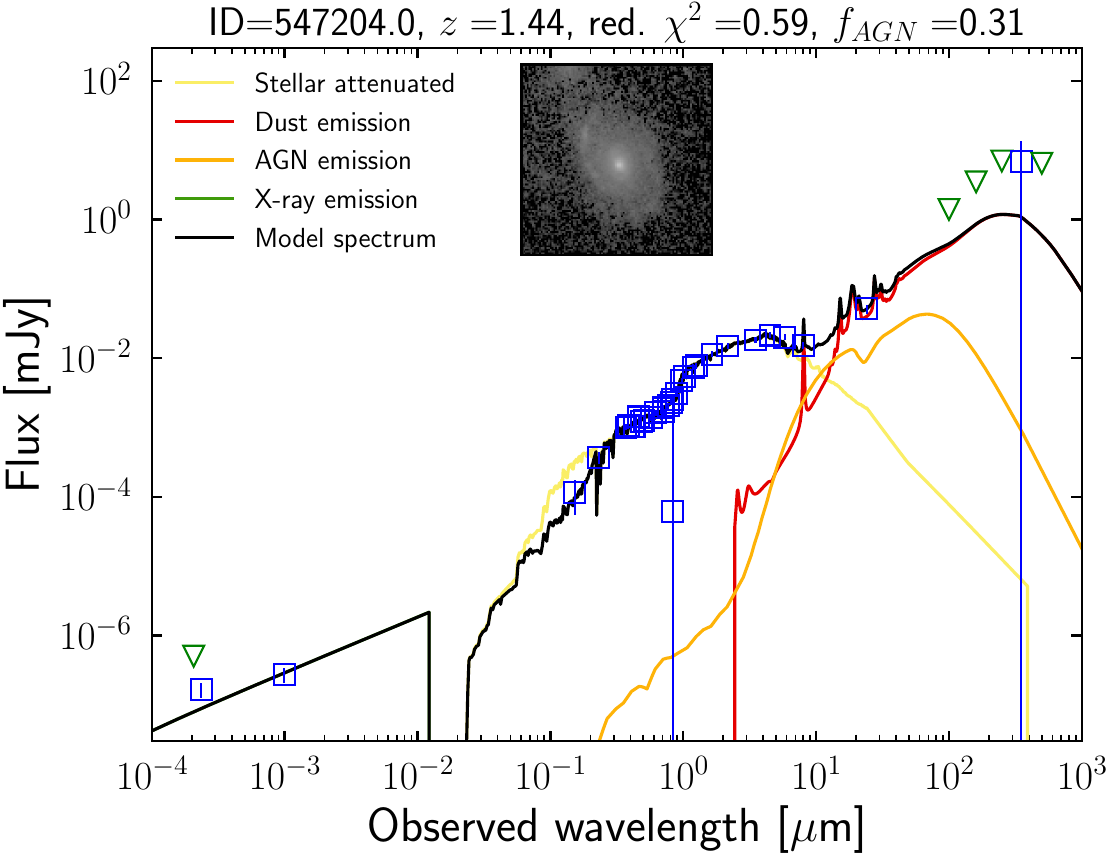}
    \includegraphics[width=.33\textwidth]{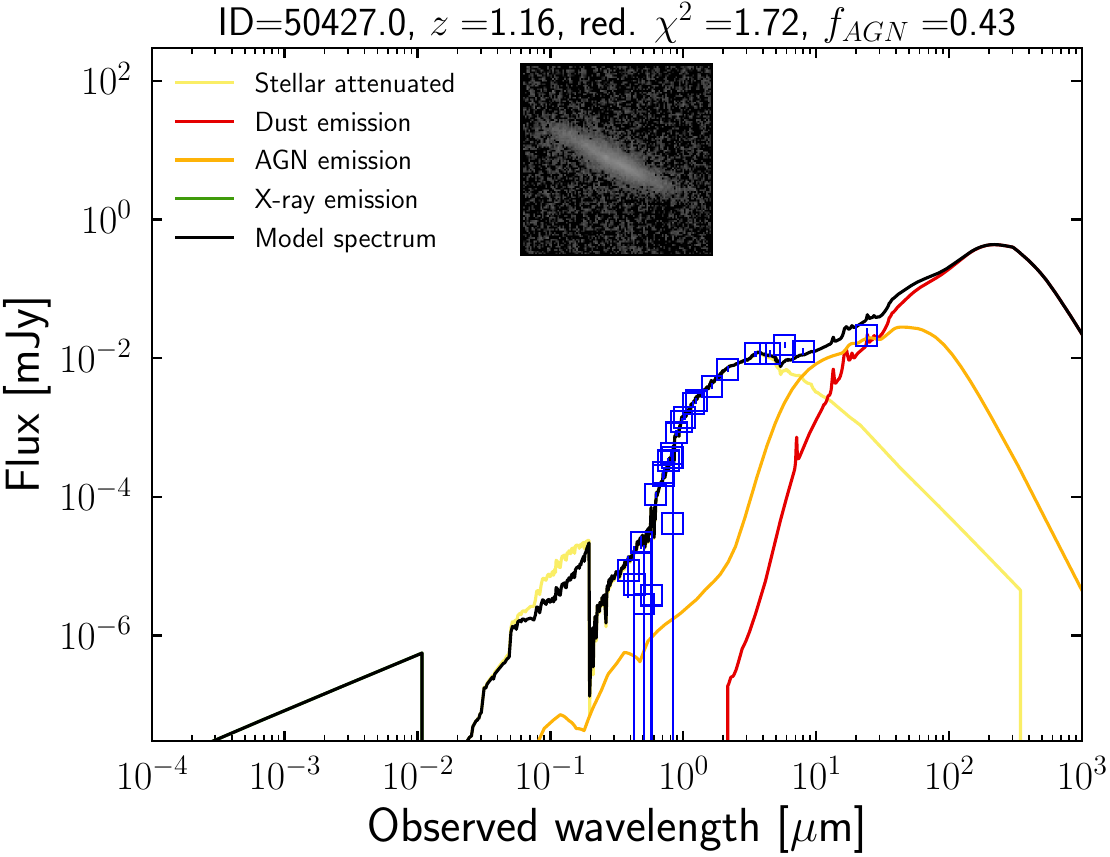}
    \includegraphics[width=.33\textwidth]{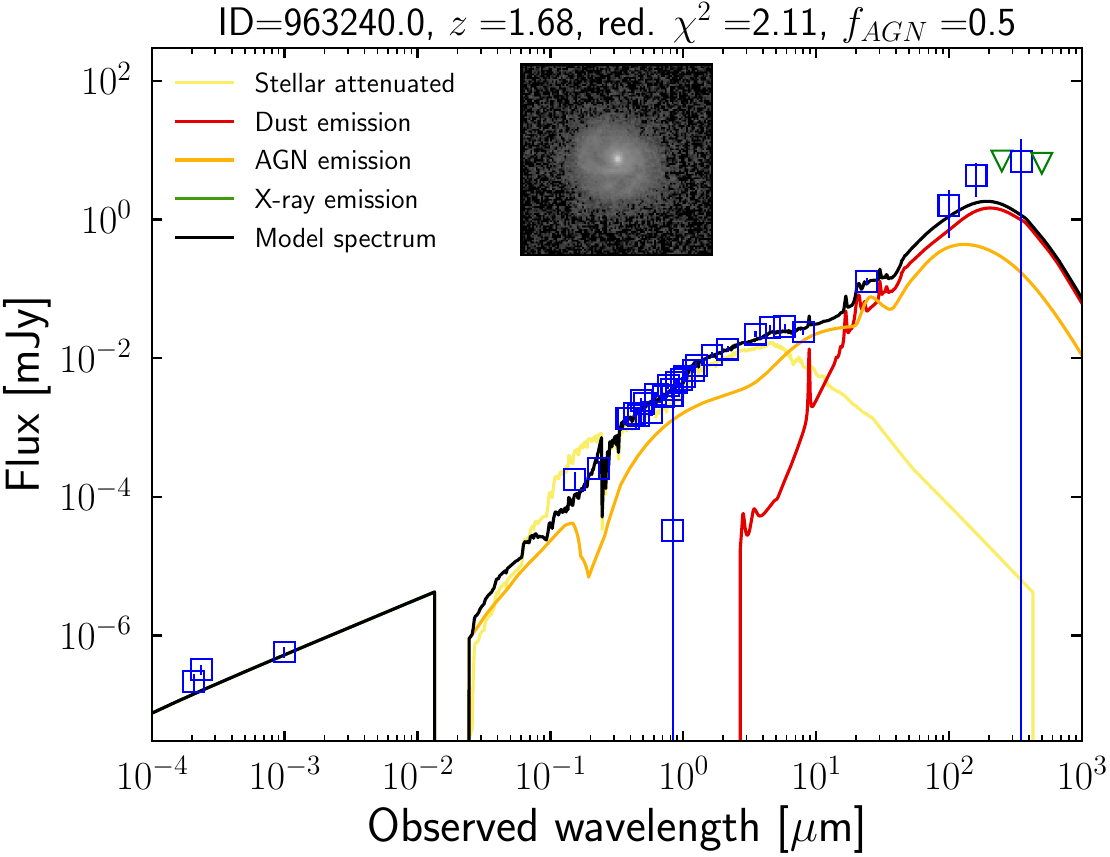}
    \includegraphics[width=.33\textwidth]{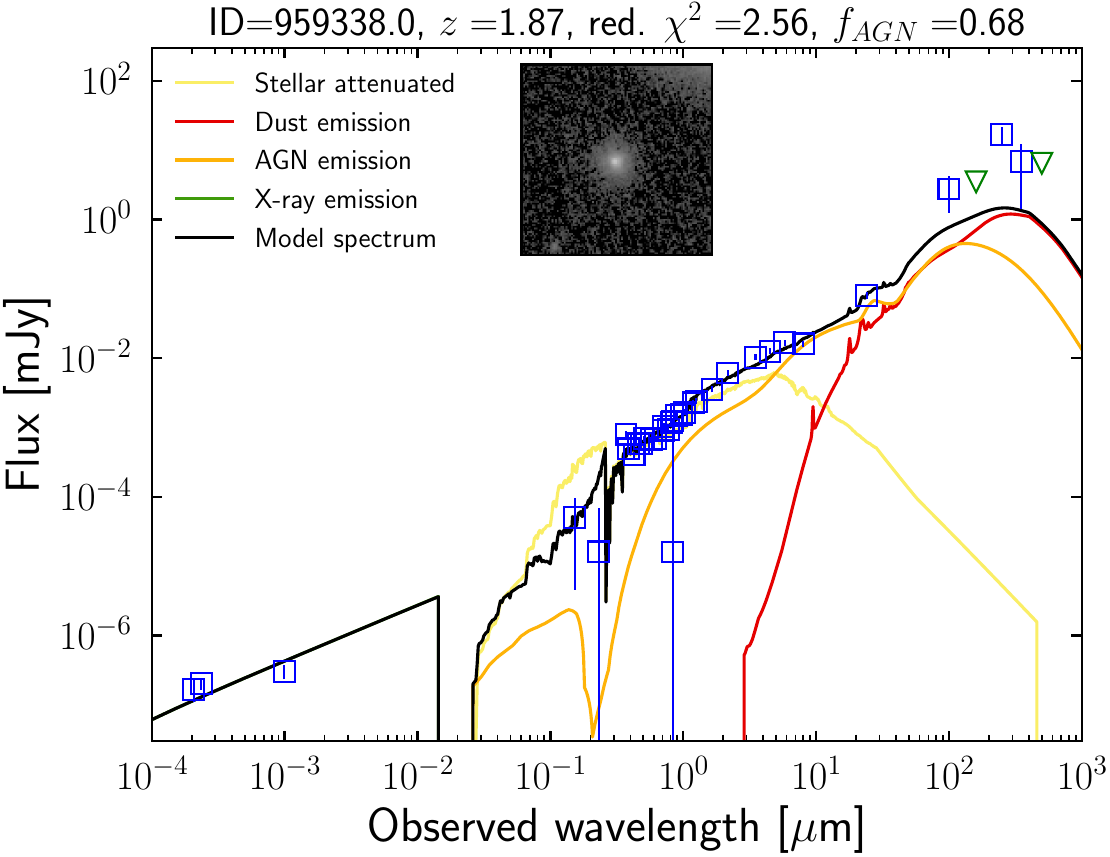}
    \includegraphics[width=.33\textwidth]{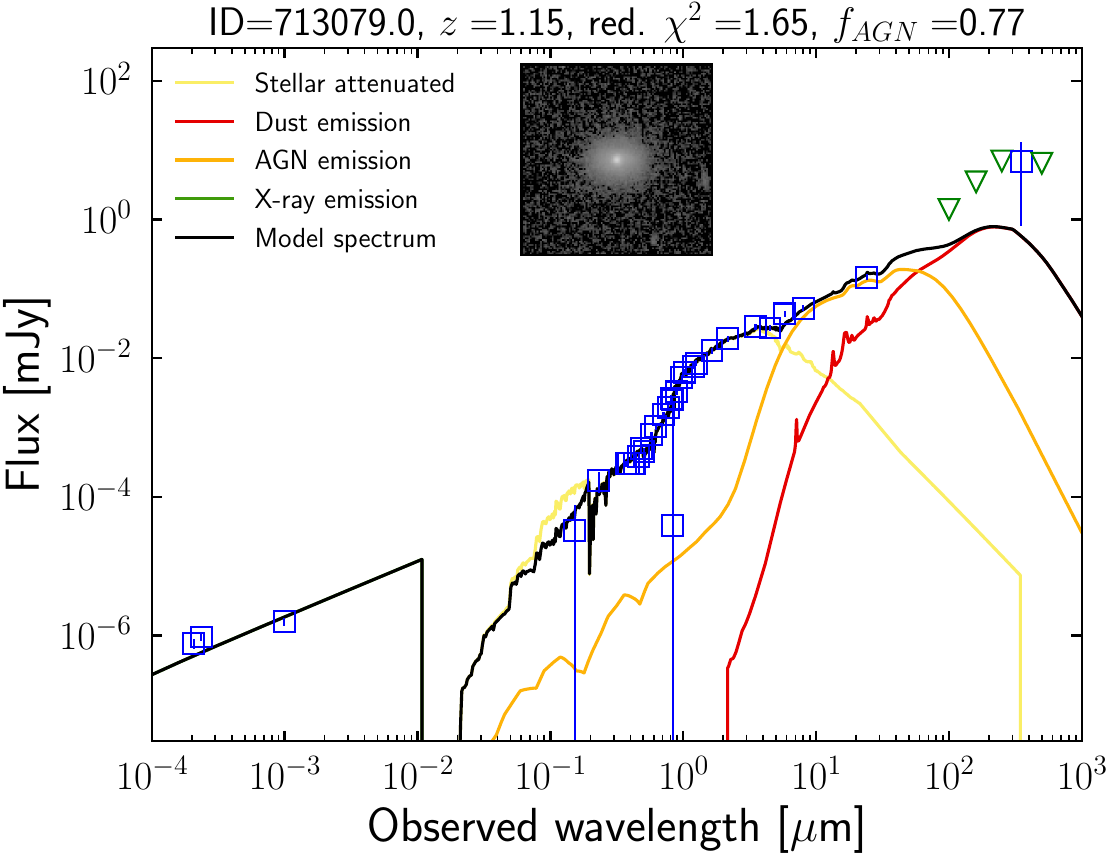}    
    \includegraphics[width=.33\textwidth]{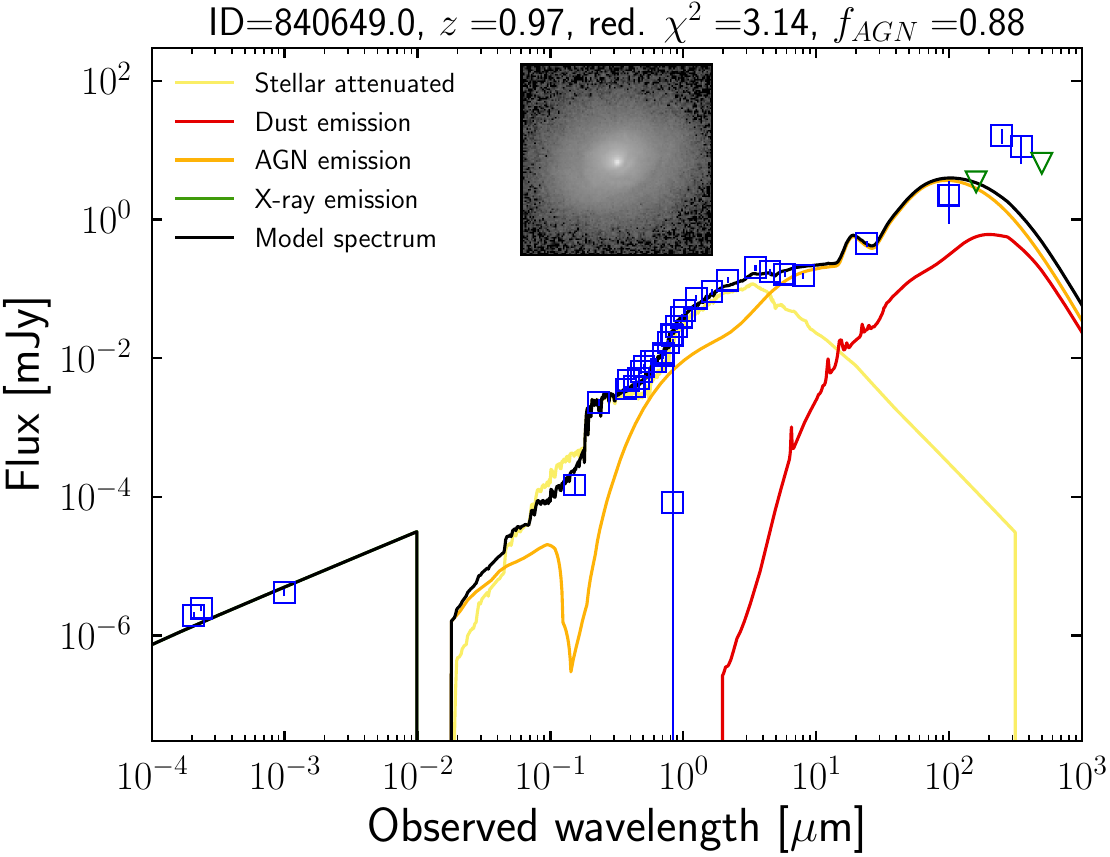}
    \includegraphics[width=.33\textwidth]{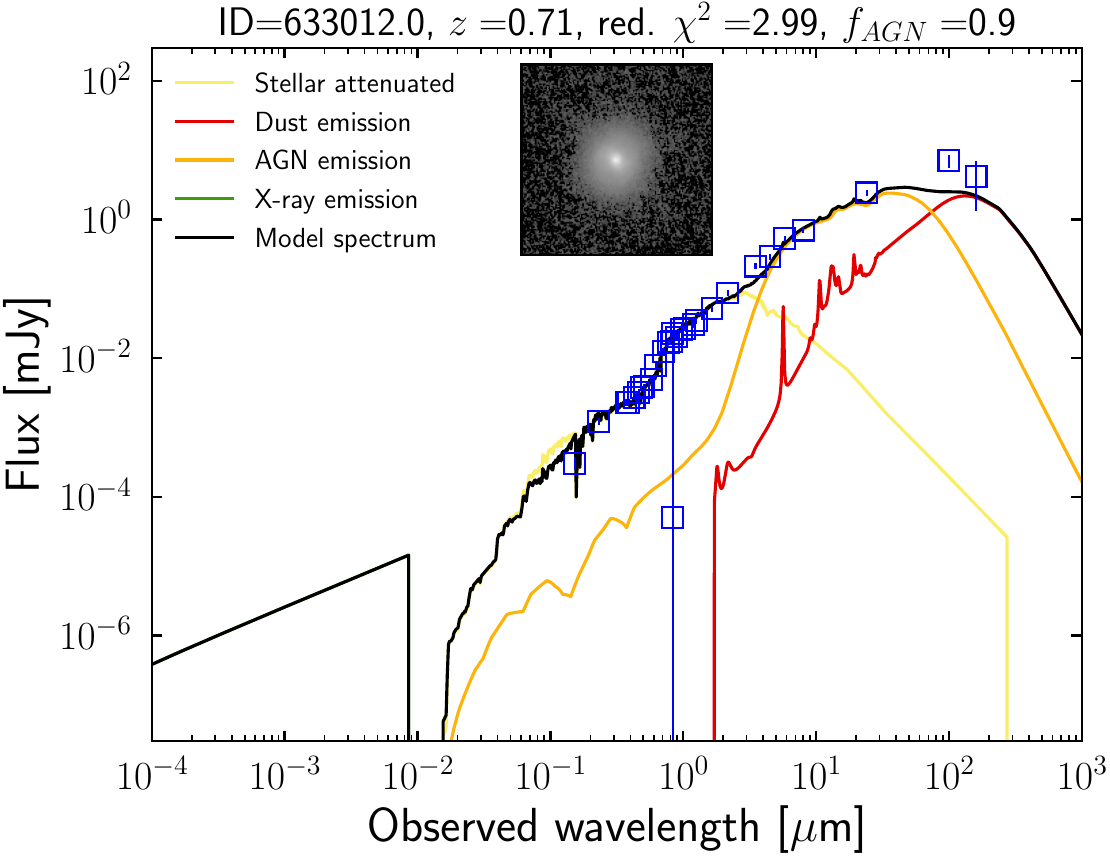}
    
    \caption{Examples of the \texttt{CIGALE} best-fit SEDs for galaxies with various levels of AGN contribution (increasing from left to right, upper to lower). The title of each panel gives the galaxy COSMOS2020 ID, redshift, reduced $\chi^2$, and the AGN fraction. Blue squares represent observed fluxes, while green triangles indicate upper limits. In each panel, we show the JWST/NIRCam F150W image of the galaxy. The cutouts have a physical size of $40\times40\,{\rm kpc}$ and are displayed using a logarithmic stretch.}
    \label{fig:sed_examples}
\end{figure*}

\begin{figure*}
    % \centering
    \sidecaption
    \includegraphics[width=.49\textwidth]{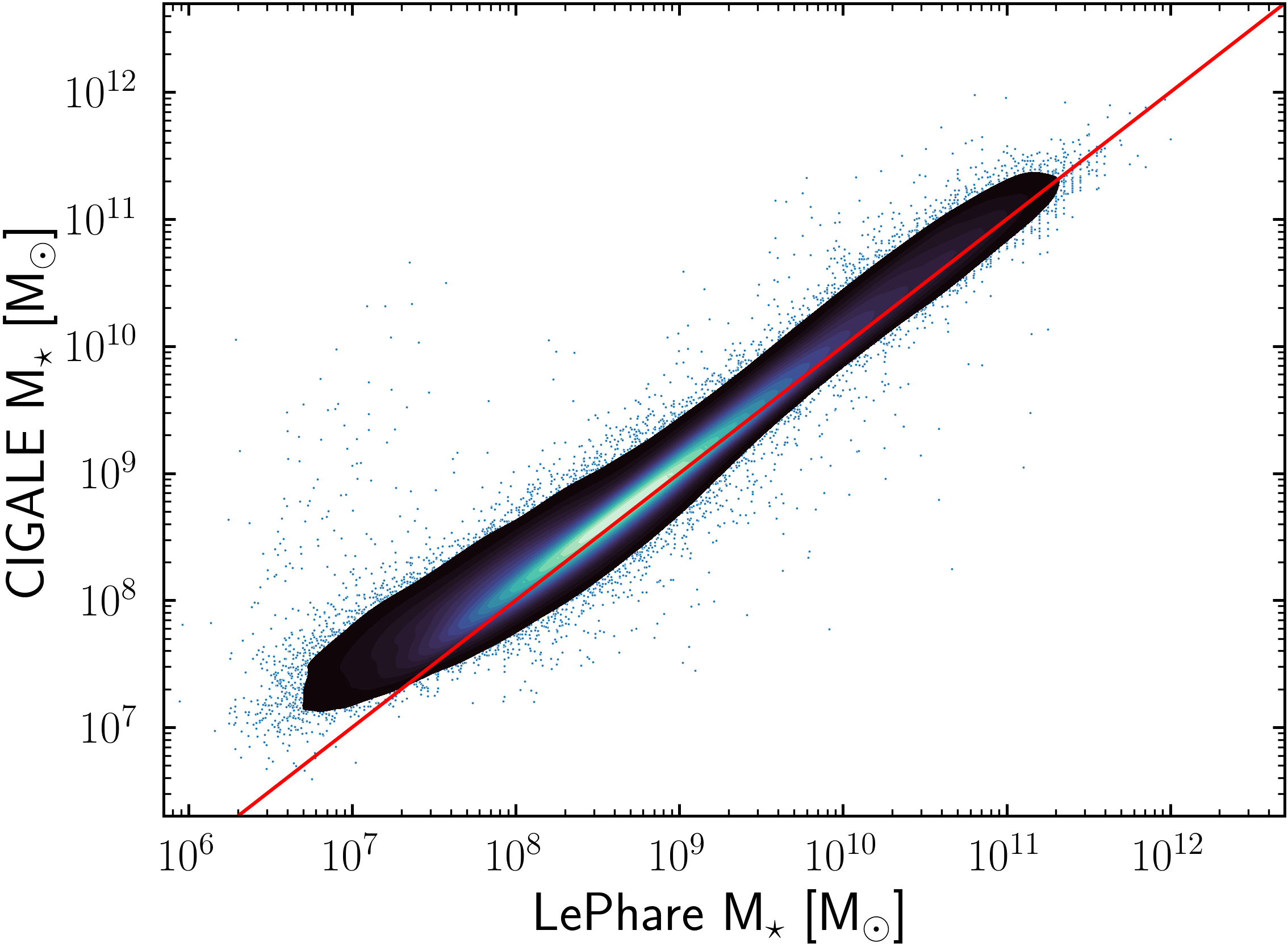}
    \caption{Comparison of the stellar mass measured by \texttt{CIGALE} in this work ($y-$axis) and the M$_{\star}$ provided in the COSMOS2020 catalogue, computed with \texttt{LePhare} \citep[$x-$axis,][]{weaverCOSMOS2020PanchromaticView2022}. 
    }
    \label{fig:Mstar_comp}
\end{figure*}

\texttt{CIGALE} offers the option to generate and analyse mock catalogues based on the best-fit model of each source of the dataset. When the mock analysis is performed, the code modifies each best flux by injecting Gaussian noise drawn from a distribution with the same standard deviation as the observed flux. The generated mock data were analysed in the same way as the observed data. The reliability of each estimated parameter can be examined by comparing the input and output values (the values from the best fit of the observed sample and the Bayesian values from the fit of the mock dataset, respectively). We used the mock analysis to study the precision of M$_{\star}$, SFR, and $f_{\rm AGN}$. We show this comparison in Fig.~\ref{fig:mock_analysis}, which reveals a good agreement between the ground truth and mock-estimated values for each parameter. M$_{\star}$ is recovered accurately across the whole mass range, with an increasing, but still small, scatter toward less massive galaxies. The SFRs show a similar trend, but in this case, the scatter is asymmetric, with the mock values overestimating the true SFR values. Overall, the AGN fraction is nicely recovered as well but with a larger dispersion around the 1-to-1 relation. Quantitatively, the median differences of the estimated M$_{\star}$, SFR, and $f_{\rm AGN}$ of the mock data from the ground truth values are:
\begin{align*}\label{control_eq}
    & \text{median}(log \; \text{M}_{\star,\rm true} - log \; \text{M}_{\star,\rm mock}) = -0.01\, ,\\
    & \text{median}(log \; \text{SFR}_{\rm true} - log \; \text{SFR}_{\rm mock}) = -0.02\, ,\\
    & \text{median}(f_{\rm AGN,\, true} - log \; f_{\rm AGN,\, mock}) = -0.01\, .\\
\end{align*}
The corresponding dispersions, computed as the MAD of the data, are 0.04, 0.06, and 0.09.
Given these results, we conclude that \texttt{CIGALE} successfully recovers the true M$_{\star}$, SFR, and $f_{\rm AGN}$ values of the mock sources. 
In Fig.~\ref{fig:sed_examples}, we display examples of the best-fit SEDs obtained with our \texttt{CIGALE} configuration. To illustrate the effect of different AGN fractions, we show examples with increasing $f_{\rm AGN}$, at various $z$. 

To assess the quality of our SED fitting procedure, we compare the stellar masses obtained with the M$_{\star}$ computed by the COSMOS2020 team using a different SED fitting software \citep[\texttt{LePhare};][]{weaverCOSMOS2020PanchromaticView2022}. Figure~\ref{fig:Mstar_comp} shows an excellent agreement between the two M$_{\star}$ estimates, with most galaxies on the one-to-one relation. We calculate the median bias, 
\begin{equation}
    b = \text{median}(\Delta {\rm M}_{\star}),
\end{equation}
and the median absolute deviation (MAD),
\begin{equation}
    \sigma_{MAD} = 1.48\times \text{median} (|\Delta {\rm M}_{\star} - \text{median}(\Delta {\rm M}_{\star})|)\, ,
\end{equation}
where $\Delta {\rm M}_{\star} = \log_{10}{\rm M}_{\star,LePhare} - \log_{10}{\rm M}_{\star,CIGALE}$.
We found a median bias of $-0.09$ and a MAD of $0.15$.

\section{CNN architecture}\label{app:CNN}

\begin{table}[]
\caption{CNN architecture.}
\small
\begin{tabular}{lccc}
\hline\hline\\[-7pt]
Layer type & \# Param. & Output shape & Properties \\ 
\hline\\[-7pt]
Input & 0 & (1,N,N) & \\
\hline
\begin{tabular}[l]{@{}l@{}}Convolutional\\ 32 filters (7,7)\end{tabular} & 1600 & (32,N,N) & \begin{tabular}[c]{@{}c@{}}1 pixels stride, \\ ``same'' padding, \\ ReLU act.\end{tabular} \\
Max Pooling & 0 & (32,N/2,N/2) & pool size 2 \\
Dropout & 0 & (32,N/2,N/2) & 25\% \\
\hline

\begin{tabular}[l]{@{}l@{}}Convolutional\\ 64 filters (7,7)\end{tabular} & 100416 & (64,N/2,N/2) & \begin{tabular}[c]{@{}c@{}}1 pixels stride, \\ ``same'' padding, \\ ReLU act.\end{tabular} \\
Max Pooling & 0 & (64,N/4,N/4) & pool size 2 \\
Dropout & 0 & (64,N/4,N/4) & 30\% \\
Batch Norm. & 256 & (64,N/4,N/4) & \\
\hline

\begin{tabular}[l]{@{}l@{}}Convolutional\\ 128 filters (7,7)\end{tabular} & 401536 & (128,N/4,N/4) & \begin{tabular}[c]{@{}c@{}}1 pixels stride, \\ ``same'' padding, \\ ReLU act.\end{tabular} \\
Max Pooling & 0 & (128,N/8,N/8) & pool size 2 \\
Dropout & 0 & (128,N/8,N/8) & 30\% \\
\hline

\begin{tabular}[l]{@{}l@{}}Convolutional\\ 128 filters (7,7)\end{tabular} & 802944 & (128,N/8,N/8) & \begin{tabular}[c]{@{}c@{}}1 pixels stride, \\ ``same'' padding, \\ ReLU act.\end{tabular} \\
Max Pooling & 0 & (128,N/16,N/16) & pool size 2 \\
Dropout & 0 & (128,N/16,N/16) & 30\% \\
\hline

Flatten & 0 & (N$^2/2$) & \\

Dense & \begin{tabular}[c]{@{}c@{}}(N$^2/2$+1)\\$\times 512$\end{tabular} & (512) & \begin{tabular}[c]{@{}c@{}}512 units, \\ ReLU act.\end{tabular} \\
Dropout & 0 & (512) & 35\% \\
Dense & 65664 & (128) & \begin{tabular}[c]{@{}c@{}}128 units, \\ ReLU act.\end{tabular} \\
Dropout & 0 & (128) & 35\% \\
Dense & 129 & (1) & \begin{tabular}[c]{@{}c@{}}1 unit, \\ sigmoid act.\end{tabular} \\
\hline
\end{tabular}
%\tablefoot{}
\label{tab:CNN}
\tablefoot{
Columns are the name of the Keras layer, the number of trainable parameters, output, and hyper-parameters for each layer. N is the size of the input images (256 / 64 for the original resolution / ``degraded'' images). ReLU activation function stands for Rectified Linear Unit. 
}
\end{table}

We present the CNN architecture and specific hyper-parameters adopted in this work in Table.~\ref{tab:CNN}, including filter numbers and sizes, total number of trainable parameters, dropout rates, strides, and activation functions. All hyper-parameters values were chosen based on a grid search. 

\section{Example mergers and non-mergers}\label{app:images}

\begin{figure*}
    \centering
    \includegraphics[width=.92\textwidth]{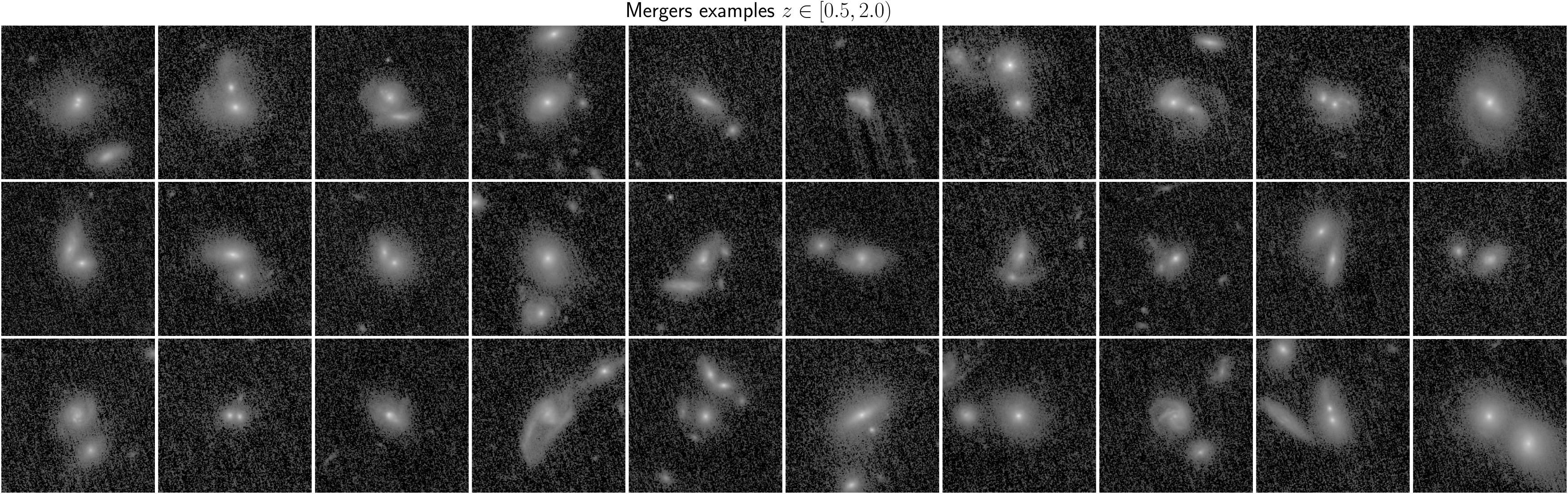}
    \includegraphics[width=.92\textwidth]{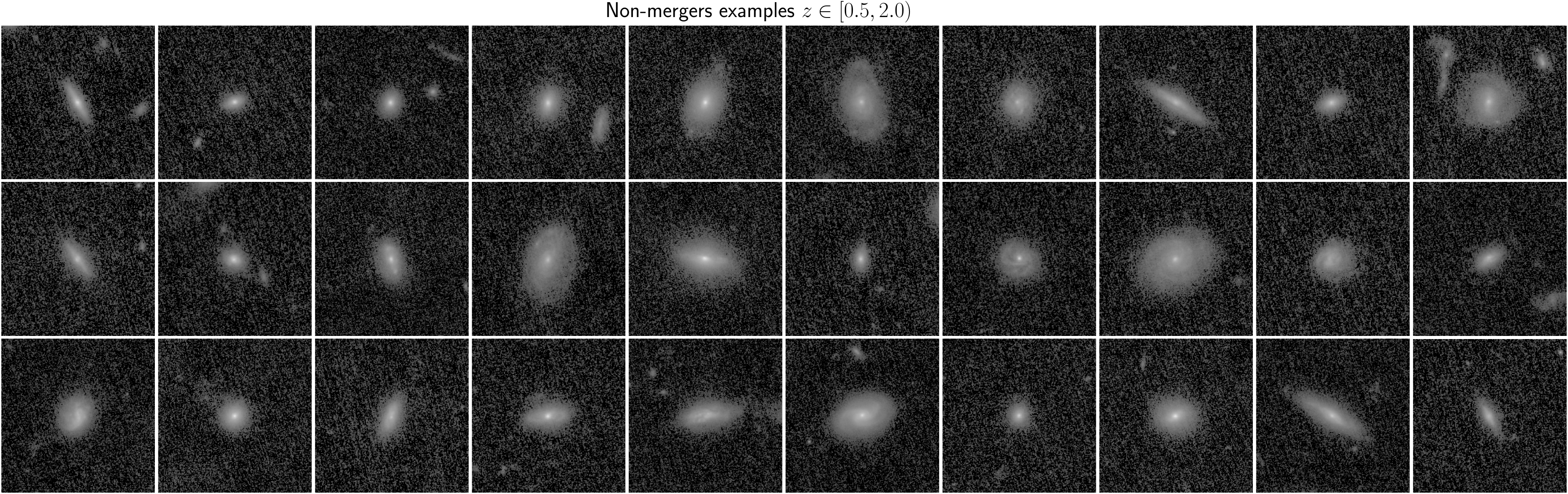}
    \caption{Random examples of galaxies classified as mergers (top 3 rows) and as non-mergers (bottom 3 rows) by our combined classifier. On top of each cutout (from the JWST/F150W images), we report the galaxy redshift and its unique ID. Cutouts have a physical size of $50\,\rm{kpc}\times50\,\rm{kpc}$, displayed using a logarithmic stretch. }
    \label{fig:img_examples_merg}
\end{figure*}

% \begin{figure*}
%     \centering
%     \includegraphics[width=.95\textwidth]{Plots_paper/JWST_z0-2_example_Non-mergers.pdf}
%     \caption{Examples of galaxies classified as non-mergers by the combined classifier. Image details as in Fig.~\ref{fig:img_examples_merg}.}
%     \label{fig:img_examples_non-merg}
% \end{figure*}

In Fig.~\ref{fig:img_examples_merg}, we show some examples of galaxies labelled as mergers and non-mergers by our classifier. Galaxies are randomly selected in the two categories. In most cases, galaxies classified as mergers appear in pairs with different degrees of morphological disturbance. Non-mergers appear as isolated and regular galaxies, with spiral arms visible in some cases.

\section{MS -- additional plots}\label{app:SFMS}

\begin{figure*}
    \centering
    \includegraphics[width=.9\textwidth]{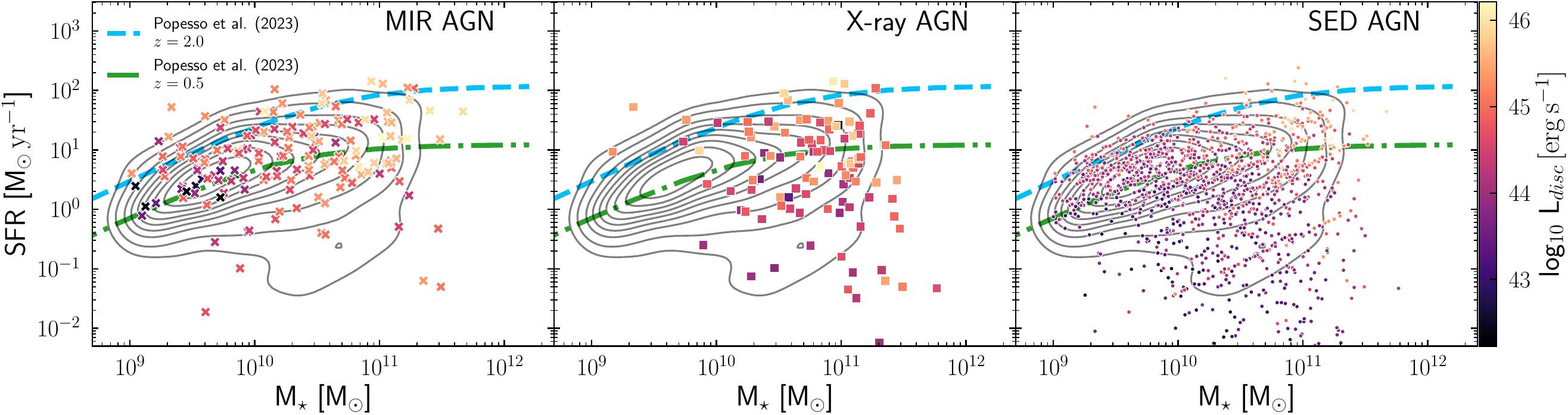}
    \caption{
    Contour plot illustrating the SFR vs. M$_{\star}$ population distribution divided by AGN type, colour-coded by the AGN disc luminosity (L$_{disc}$). In each panel, the contours (from 10\% to 90\%, with intervals of 10\%) represent the non-AGN distribution. The dashed blue and the dash-dotted green lines indicate the MS at $z=2$ and $z=0.5$ from \citet{popessoMainSequenceStarforming2023}, respectively.}
    \label{fig:AGN_MS_LAGN}
\end{figure*}

\begin{figure*}
    % \centering
    \sidecaption
    \includegraphics[width=.45\textwidth]{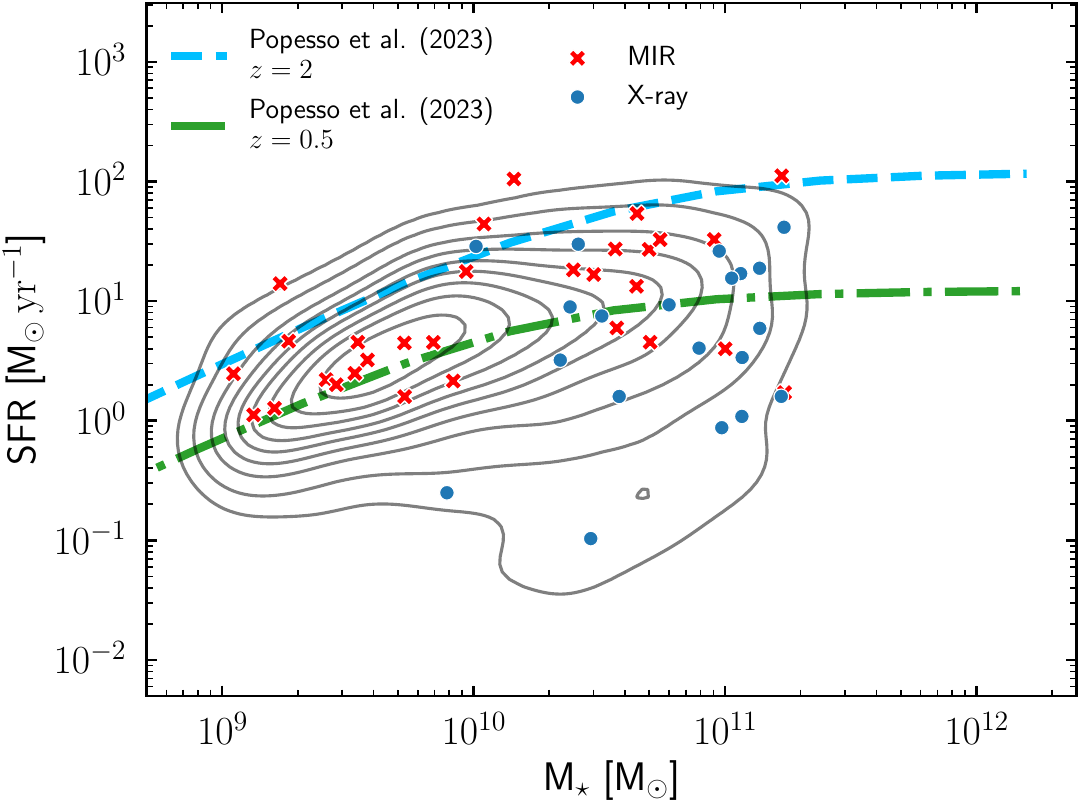}
    \caption{
    SFR vs. M$_{\star}$ distribution, but in this case, both MIR and X-ray AGN types do not include the other AGN types. %The dashed blue and the dash-dotted green lines indicate the MS at $z=2$ and $z=0.5$ from \citet{popesso_main_2023}, respectively. Contours represent the same levels as in Fig.~\ref{fig:AGN_MS}.
    }
    \label{fig:AGN_MS_excl}
\end{figure*}

We further explored the SFR vs. stellar mass relation for AGN and non-AGN host galaxies. 
In Fig.~\ref{fig:AGN_MS_LAGN}, we plot this relation for each AGN type, colour-coded by AGN accretion disc luminosity (L$_{disc}$). The most luminous AGN are more likely to be hosted by star-forming galaxies than by low SFR ones, independently of AGN type. A slight correlation also appears with M$_{\star}$: the most luminous AGN reside in the most massive galaxies.
In Fig.~\ref{fig:AGN_MS_excl}, we show the same relation, but for MIR and X-ray AGN that are exclusively in that specific AGN class, i.e. we do not include AGN identified with multiple methods. 
Once again, X-ray AGN reside mostly in massive galaxies on or below the MS, while MIR AGN are in relatively less massive galaxies on or above the MS. 
% Non-AGN and SED AGN are dominated by low-mass galaxies and generally follow the MS, but we see non-negligible fractions of galaxies in the starburst sequence and the quenched area. 

Overall, these trends reflect the general trend of AGN and non-AGN when all redshifts are combined and when overlapping between AGN types is allowed (Fig.~\ref{fig:AGN_MS}).
Therefore, we can conclude that: X-ray AGN are preferably hosted by giant, quenched or on the way to be quenched galaxies; MIR AGN are usually in intermediate/high M$_{\star}$ star-forming galaxies; SED AGN and non-AGN mostly follow the MS, but some are also found in the starburst and quenched regions.

\section{Binary experiments results}\label{app:tables}

\begin{table}%[h]
\caption{Frequency of MIR-, X-ray-, and SED-selected AGN in mergers (M) and non-mergers (NM) in different redshift bins. }
\small
% \footnotesize
  \centering
  \begin{tabular}{lccc}
  \hline\hline\\[-7pt]
  & \multicolumn{3}{c}{MIR AGN} \\
  $z$ bin & M & NM (ctrl.) & Excess \\
  \hline\\[-7pt]
  0.5--1.25 & $1.4\pm 0.4 \%$ & $1.1 \pm 0.1 \%$ & $1.3\pm0.4$ \\
  % & & 
  & (14/981) & (111/9887) & \\
  \hline\\[-7pt]
  1.25--2.0 & $4.0\pm0.6 \%$ & $2.7\pm0.2 \%$ & $1.4\pm0.2$ \\
  % & 
  & (43/1086) & (296/10783) & \\
  % & (490/1086) & (4332/10783) & \\
  
  \hline\\[-7pt]
  & \multicolumn{3}{c}{X-ray AGN} \\
  $z$ bin & M & NM (ctrl.) & Excess \\
  \hline\\[-7pt]
  0.5--1.25 & $1.9\pm 0.4 \%$ & $0.7 \pm 0.1 \%$ & $2.6\pm0.6$ \\
  & (19/981) & (74/9887) & \\
  \hline\\[-7pt]
  1.25--2.0 & $2.2 \pm 0.5 \%$ & $0.87 \pm 0.08 \%$ & $2.6\pm 0.6$ \\
  & (24/1086) & (92/10783) & \\

  \hline\\[-7pt]
  & \multicolumn{3}{c}{SED AGN} \\
  $z$ bin & M & NM (ctrl.) & Excess \\
  \hline\\[-7pt]
  0.5--1.25 & $14\pm 1\%$ & $12.5 \pm 0.3 \%$ & $1.10\pm 0.09$ \\ 
  & (135/981) & (1234/9887) & \\
  \hline\\[-7pt]
  1.25--2.0 & $22 \pm 1 \%$ & $17.7 \pm 0.4 \%$ & $1.25\pm0.08$ \\
  & (241/1086) & (1909/10783) & \\
  \hline
  \end{tabular}
  \label{tab:agn_freq}
  \tablefoot{
  Fractions and relative errors are calculated using bootstrapping with resampling (1000 samples for each population). In brackets, we provide the number of AGN for each type over the total number of mergers and non-merger controls in each $z$-bin. 
}
\end{table}

% \begin{table*}[h]
% \caption{Merger fraction ($f_{merg}$) in galaxies hosting MIR-, X-ray-, and SED-selected AGN, and the $f_{merg}$ relative to non-AGN controls. }
% \small
%   \centering
%   \begin{tabular}{lcccccc}
%   \hline\hline\\[-7pt]
%   & \multicolumn{2}{c}{$f_{merg}$ in} & \multicolumn{2}{c}{$f_{merg}$ in} & \multicolumn{2}{c}{$f_{merg}$ in} \\
%   $z$ bin & MIR AGN & Non-AGN ctrl. & X-ray AGN & Non-AGN ctrl. & SED AGN & Non-AGN ctrl. \\
%   \hline\\[-7pt]
%   0.5--1.25 & $27\pm6\%$ & $17.8\pm1.6\%$ & $48\pm7\%$ & $22\pm2\%$ & $24.7\pm1.4\%$ & $18.4\pm0.4\%$ \\
%   & (16/59) & (106/597) & (22/46) & (103/472) & (251/1019) & (1941/10537) \\
%   \hline\\[-7pt]
%   1.25--2.0 & $48\pm5\%$ & $36\pm2\%$ & $56\pm9\%$ & $28\pm2\%$ & $39.1\pm1.5\%$ & $34.8\pm0.5\%$ \\
%   & (40/83) & (300/823) & (19/34) & (93/328) & (398/1021) & (3433/9863) \\
%   \hline
%   \end{tabular}
%   \label{tab:merg_frac}
%   \tablefoot{
%   Fractions and relative errors are calculated using bootstrapping with resampling (1000 samples for each population). In brackets, we provide the numbers of AGN for each type over the total number of mergers and non-merger controls in each $z$-bin. 
% }
% \end{table*}

\begin{table}[h]
\caption{Merger fraction ($f_{\rm merg}$) in galaxies hosting MIR-, X-ray-, and SED-selected AGN, and the $f_{\rm merg}$ relative to non-AGN controls. }
\small
  \centering
  \begin{tabular}{lcc}
  \hline\hline\\[-7pt]
  & \multicolumn{2}{c}{$f_{\rm merg}$ in} \\ 
  $z$ bin & MIR AGN & Non-AGN ctrl. \\
  \hline\\[-7pt]
  0.5--1.25 & $27\pm6\%$ & $17.8\pm1.6\%$ \\
  & (16/59) & (106/597) \\
  \hline\\[-7pt]
  1.25--2.0 & $48\pm5\%$ & $36\pm2\%$ \\ 
  & (40/83) & (300/823) \\
  
  \hline\\[-7pt]
  & \multicolumn{2}{c}{$f_{\rm merg}$ in} \\ 
  $z$ bin & X-ray AGN & Non-AGN ctrl. \\
  \hline\\[-7pt]
  0.5--1.25 & $48\pm7\%$ & $22\pm2\%$ \\
  & (22/46) & (103/472) \\
  \hline\\[-7pt]
  1.25--2.0 & $56\pm9\%$ & $28\pm2\%$ \\
  & (19/34) & (93/328) \\
  
  \hline\\[-7pt]
  & \multicolumn{2}{c}{$f_{\rm merg}$ in} \\ 
  $z$ bin & SED AGN & Non-AGN ctrl. \\
  \hline\\[-7pt]
  0.5--1.25 & $23\pm 2\%$ & $17.0\pm0.5\%$ \\
  & (128/567) & (995/5793) \\
  \hline\\[-7pt]
  1.25--2.0 & $40\pm2\%$ & $34.0\pm0.7\%$ \\
  & (192/482) & (1598/4697) \\
  \hline
  \end{tabular}
  \label{tab:merg_frac}
  \tablefoot{
  Fractions and relative errors are calculated using bootstrapping with resampling (1000 samples for each population). In brackets, we provide the number of AGN for each type over the total number of mergers and non-merger controls in each $z$-bin. 
}
\end{table}

In this appendix section, we report the numerical results of the merger--AGN relation adopting a binary AGN/non-AGN classification. Table~\ref{tab:agn_freq} shows the AGN frequency measured in mergers and non-mergers for all AGN types, divided into two redshift bins (results presented in Fig.~\ref{fig:agn_freq}). Table~\ref{tab:merg_frac} provides the merger fractions for all AGN types and relative non-AGN controls in the two redshift bins, as presented in Fig.~\ref{fig:merg_frac}.

\section{$f_{\rm merg}$--$f_{\rm AGN}$ parametrisation}\label{app:fit}

In this section, we detail the fitting procedure adopted to parametrise the $f_{\rm merg}$--$f_{\rm AGN}$ relation presented in \citetalias{lamarcaDustPowerUnravelling2024a}. In the case of MIR and X-ray AGN, we use a power law to fit the data, in the form:
\begin{equation}
    y = x^{\alpha} + c\, ,
    \label{eq:power_law}
\end{equation}
while a combination of two power laws is used for the SED AGN,
\begin{equation}
    y = x^{\alpha} + \beta \cdot x^{\gamma}\, ,
    \label{eq:double_power_law}
\end{equation}
where $x = f_{\rm AGN}$ and $y=f_{\rm merg}$. To enhance accuracy in the fits, we incorporate a sampling method. First, we uniformly sample the number of $f_{\rm AGN}$ bins to divide the sample. The number of bins is drawn in the range of $6-20$ for the X-ray and SED AGN, while the maximum number of bins allowed for the MIR AGN is set to 15\footnote{The lower number statistics of MIR AGNs mainly led to this choice.}. We calculate the $f_{\rm merg}$ and its uncertainty for each bin. To account for these uncertainties, which we assume follow a Gaussian distribution, we apply a bootstrapping method that varies the fractions within their errors. Next, we find the best-fit parameters for the data and repeat this process 10,000 times to ensure robustness. Finally, we calculate the best parameters as the median values of the 10,000 best fits. The best-fit parameters are reported in Table~\ref{tab:best_fit}.

\begin{table}
    \centering
    \small
    \caption{The best-fit parameters of Eq.~\ref{eq:power_law} for X-ray and MIR AGNs, and of Eq.~\ref{eq:double_power_law} for SED AGNs.}
    \begin{tabular}{lcc}
    \hline\hline\\[-7pt]
    AGN type & $\alpha$ & $c$ \\
    \hline\\[-7pt]
    \noalign{\vskip 2mm}
    X-ray & $16.2^{+4.87}_{-3.52}$ & $0.287^{+0.009}_{-0.009}$ \\
    \noalign{\vskip 2mm}
    MIR & $35.0^{+43.1}_{-9.46}$ & $0.478^{+0.022}_{-0.023}$ \\
    \noalign{\vskip 2mm}
    \hline
    \vspace{2mm}
    \end{tabular}
    
    \begin{tabular}{lccc}
    \hline\hline\\[-7pt]
    AGN type & $\alpha$ & $\beta$ & $\gamma$ \\
    \hline\\[-7pt]
    \noalign{\vskip 2mm}
    SED & $47.2^{+29.1}_{-14.0}$ & $0.432^{+0.019}_{-0.018}$ & $0.637^{+0.041}_{-0.030}$ \\
    \noalign{\vskip 2mm}
    \hline
    \end{tabular}
    \tablefoot{Each parameter is estimated using a resampling process. The values reported are the median values of the 10,000 best-fit parameters and their $25^{th} - 75^{th}$ percentile ranges. }
    \label{tab:best_fit}
\end{table}

\end{appendix}

%--------------------------------------------------------------------

\end{document}